\documentclass[prd,superscriptaddress,amsfonts,amssymb,amsmath,showpacs,twocolumn,nofootinbib]{revtex4-2}
\usepackage{bm}
\usepackage{amsfonts}
\usepackage{latexsym}
\usepackage{graphicx}
\usepackage{amsmath}
\usepackage{palatino}
\usepackage{xcolor} 
\usepackage{mathpazo}
\usepackage{xcolor}
\usepackage{rotating}
\usepackage{adjustbox}
\usepackage{tensor}
\usepackage{textcomp}
\usepackage{float}
\usepackage{booktabs}
\usepackage{dcolumn}
\usepackage[titletoc]{appendix}
\usepackage{booktabs}
\usepackage{multirow}
\usepackage{hyperref}
\hypersetup{colorlinks,citecolor=blue}
\usepackage{amsmath}
\usepackage{xcolor}
\usepackage{orcidlink}
\usepackage{epsfig}
\usepackage{caption}
\usepackage{subcaption}
\usepackage{commath}
\hypersetup{colorlinks,citecolor=blue}
\hypersetup{colorlinks=true,linkcolor=magenta,filecolor=magenta,    urlcolor=blue}

\def\be{\begin{equation}}
\def\ee{\end{equation}}
\def\bea{\begin{eqnarray}}
\def\eea{\end{eqnarray}}

\definecolor{owngreen}{rgb}{0.0, 0.5, 0.0}

\begin{document}

\title{Is the $\Lambda$CDM Model in Crisis?}

\author{Himanshu Chaudhary}
\email{himanshu.chaudhary@ubbcluj.ro,\\
himanshuch1729@gmail.com}
\affiliation{Department of Physics, Babeș-Bolyai University, Kogălniceanu Street, Cluj-Napoca, 400084, Romania}
\author{Salvatore Capozziello}
\email{capozziello@na.infn.it}
\affiliation{Dipartimento di Fisica ``E. Pancini", Universit\`a di Napoli ``Federico II", Complesso Universitario di Monte Sant’ Angelo, Edificio G, Via Cinthia, I-80126, Napoli, Italy,}
\affiliation{Istituto Nazionale di Fisica Nucleare (INFN), sez. di Napoli, Via Cinthia 9, I-80126 Napoli, Italy,}
\affiliation{Scuola Superiore Meridionale, Largo S. Marcellino, I-80138, Napoli, Italy.}
\author{Subhrat Praharaj}
\email{s.praharaj@uva.nl,\\
sp7437473@@gmail.com}
\affiliation{Anton Pannekoek Instituut, University of Amsterdam, Science Park 904, 1098 XH Amsterdam}
\author{Shibesh Kumar Jas Pacif}
\email{shibesh.math@gmail.com}
\affiliation{Pacif Institute of Cosmology and Selfology (PICS), Sagara, Sambalpur 768224, Odisha, India}
\affiliation{Research Center of Astrophysics and Cosmology, Khazar University, Baku, AZ1096, 41 Mehseti Street, Azerbaijan}
\author{G. Mustafa}
\email{gmustafa3828@gmail.com}
\affiliation{Department of Physics,
Zhejiang Normal University, Jinhua 321004, People’s Republic of China}

\begin{abstract}
We present strong evidences for dynamical dark energy challenging the $\Lambda$CDM model. Several dark energy models are explored, including $\omega_0\omega_a$CDM, logarithmic, Exponential, JBP, and BA, along with non-flat cosmological models  accounting also for potential spatial curvature different from zero. Through our analysis, we find evidences supporting a flat Universe ($\Omega_k \approx 0$). Using the Metropolis-Hastings Markov Chain Monte Carlo algorithm, we analyze observational data from Baryon Acoustic Oscillations of DESI DR2, Type Ia Supernovae, and Compressed CMB likelihood to constrain the parameters of these models. Our findings provide strong evidences that $\omega \neq -1$, with deviations from the $\Lambda$CDM model favoring dynamical dark energy models characterized by the Quintom-B scenario ($\omega_0 > -1$, $\omega_a < 0$, and $\omega_0 + \omega_a < -1$). We also derive the upper bounds on $\sum m_\nu$ using the combination of CMB and DESI DR2 data. For the $\Lambda$CDM model, we find $\sum m_\nu < 0.066~\text{eV}$, while for $\omega$CDM, it is $\sum m_\nu < 0.075~\text{eV}$. In the o$\Lambda$CDM and o$\omega$CDM models, the limits are $\sum m_\nu < 0.263~\text{eV}$ and $\sum m_\nu < 0.520~\text{eV}$, respectively. For other models, including $\omega_0\omega_a$CDM, Logarithmic, Exponential, JBP, BA, and GEDE, the upper limits range from $< 0.043~\text{eV}$ to $< 0.127~\text{eV}$, depending on the model. Constraints on the effective number of relativistic species, $N_{\text{eff}}$, show that our results remain consistent with the standard value of $N_{\text{eff}} = 3.044$ for each dark energy model. Bayesian evidence shows that combining the DES-SN5Y and Union3 SNe Ia samples with CMB + DESI DR2 reveals a deviation from the $\Lambda$CDM model. Finally, we found that none of the models reach the $5\sigma$ threshold of deviation from $\Lambda$CDM, but some models show tensions exceeding $3\sigma$ with DES-SN5YR or Union3, indicating we are starting to see the cracks in the cosmological constant $\Lambda$.
\end{abstract}

\maketitle


\section{Introduction}\label{sec_1}
The cosmological constant ($\Lambda$) has  a long and unsettled history in cosmology. It was originally introduced by Einstein in 1917 to modify the field equations of General Relativity and obtain a static Universe \cite{einstein1917}. After the discovery of cosmic expansion, $\Lambda$ was abandoned for several decades, until the late 1990s, when the observations of the accelerated expansion of the Universe through high-redshift Type Ia supernovae \cite{riess1998, perlmutter1999} revived it as a central parameter in modern cosmology. In the concordance $\Lambda$ Cold Dark Matter ($\Lambda$CDM) model, $\Lambda$ is interpreted as a constant vacuum energy density with an equation of state $w=-1$. This model successfully fits a broad set of observations, including cosmic microwave background (CMB) anisotropies \cite{aghanim2020planck,choi2020atacama,hinshaw2013nine} baryon acoustic oscillations (BAO) \cite{eisenstein2005, DESI:2023ytc, DESI:2025zgx}, supernova luminosity distances \cite{riess1998, perlmutter1999}, large-scale structure surveys \cite{tegmark2004}, and weak lensing data \cite{desy3}.

Despite its empirical success, the cosmological constant remains theoretically puzzling. The most severe difficulty is the \textit{cosmological constant problem}, arising from the enormous discrepancy (55–120 orders of magnitude) between the observed value of $\Lambda$ and  predictions coming  from quantum field theory \cite{weinberg1989, martin2012}. A related puzzle is the \textit{cosmic coincidence problem}, which asks why the densities of matter and dark energy (DE) are of comparable magnitude today despite evolving very differently with cosmic expansion \cite{steinhardt2003}. These challenges  motivated the development of numerous alternative models.

Several approaches consider $\Lambda$ to be dynamical rather than constant. Variable$\Lambda$ cosmologies, studied extensively by Vishwakarma \cite{vishwakarma1999, vishwakarma2000}, attempt to link its evolution with the expansion history. Scalar-field DE models such as quintessence, k-essence, and phantom energy \cite{ratra1988, caldwell2002, copeland2006} introduce a time-dependent equation of state $w(z)$. Tracker fields \cite{wang2000} and coupled DE scenarios \cite{amendola2000, kumar2016} aim to alleviate the coincidence problem. Modified gravity theories, including $f(R)$ gravity \cite{Capozziello:2002rd,Capozziello:2011et, defelice2010} and scalar tensor frameworks \cite{Bamba:2012cp, clifton2012}, reinterpret cosmic acceleration without invoking a true vacuum energy. Padmanabhan \cite{padmanabhan2003, padmanabhan2008, padmanabhan2016} has argued that the effective cosmological constant may originate from vacuum fluctuations, while Sahni and collaborators \cite{sahni2000, sahni2004}  explored a wide landscape of dynamical DE models and alternative paradigms. Cooperstock and colleagues \cite{cooperstock1995, cooperstock1998} examined the role of $\Lambda$ in the energy-momentum structure of General Relativity, raising the question of its significance beyond cosmological scales. While $\Lambda$CDM remains the simplest model, \textit{observational tensions are accumulating}. The most prominent is the Hubble tension, a discrepancy $\geq 5\sigma$ between the Hubble constant $H_0$ inferred from CMB+$\Lambda$CDM analyses ($\sim 67$ km/s/Mpc) and local measurements ($\sim 73$ km/s/Mpc) \cite{freedman2021, riess2022,Abdalla:2022yfr,CosmoVerseNetwork:2025alb}.

More recently, the  baryon acoustic oscillation (BAO) dataset from the DE Spectroscopic Instrument (DESI) has provided new insights into the understanding of DE.  DESI DR1 and DR2 analyzes suggest significant deviations from $\Lambda$CDM, with indications that dynamic DE is favored over a cosmological constant. In 2024, building on indications of dynamical DE from the Pantheon+ and Union3 datasets, DESY5 found that whether using supernova data alone or in combination with CMB, BAO, and 3 × 2pt measurements, the best fit equation of the state parameter ($\omega$) consistently lies slightly above $-1$ at more than 1$\sigma$ level. These results align with findings from Union3, reinforcing the trend towards mildly dynamical DE. Furthermore, DESI's first-year BAO measurements revealed that the constraints on $\omega_0$ and $\omega_a$ deviate from $\Lambda$CDM by 2.6$\sigma$, 2.5$\sigma$, 3.5$\sigma$ and 3.9$\sigma$ when combined with data from CMB, Pantheon$^+$, Union3 and DESY5, respectively \cite{adame2025desi}. These findings favor a dynamic DE model, particularly a Quintom-B scenario characterized by $\omega_0 > -1$, $\omega_a < 0$, and $\omega_0 + \omega_a < -1$ \cite{cai2025quintom, ye2025hints,chaudhary2025evidence}. Furthermore, DESI DR2 BAO measurements, when combined with CMB, exclude $\Lambda$CDM at a 3.1$\sigma$ significance, with stronger exclusions of 2.8$\sigma$, 3.8$\sigma$, and 4.2$\sigma$ when adding Pantheon$^+$, Union3 and DESY5 data. Compared to DR1, DR2 offers improved precision and reduced uncertainties \cite{karim2025desi}. An extended analysis following the release of DR2 further confirms the presence of dynamical DE \cite{Odintsov:2024woi,lodha2025extended}. The results highlight that we should reconsider our understanding of DE. Researchers are eager to uncover the potential new physics suggested by the recent DESI measurements and to explore whether new physics in the DE sector can help resolve cosmological tensions. In this context, numerous recent studies have investigated these possibilities\cite{wolf2025robustness,shlivko2025optimal,giare2024robust,payeur2025observations,xu2025constranits,alam2025beyond,paliathanasis2025observational,paliathanasis2025challenging,paliathanasis2025dark,van2025compartmentalization,dinda2025calibration,liu2025torsion,choudhury2024updated,choudhury2025cosmology,choudhury2025cosmological,lee2025constraining,vagnozzi2020new,jiang2024nonparametric,pedrotti2025bao,fazzari2025cosmographic}.\\\\

Another critical frontier involves neutrino physics. A generic prediction of the hot Big Bang model is the cosmic neutrino background, relic neutrinos that behave as radiation in the early universe and as matter at late times. They influence both acoustic oscillations in the primordial plasma and structure formation. Cosmological observations are therefore sensitive to the \textit{sum of neutrino masses} $\Sigma m_\nu$ and the \textit{effective number of relativistic species} $N_{\rm eff}$ \cite{lesgourgues2006}. Current limits, strengthened by DESI in combination with Planck and supernova data, constrain $\Sigma m_\nu \lesssim 0.06$–0.1 eV \cite{desi2025a,desi2025b}, while deviations of $N_{\rm eff}$ from the standard value of 3.044 \cite{grohs2016} would imply additional light relics. The combination of DESI and CMB data has provided more stringent constraints on neutrino masses. For DESI DR2, the 95\% upper limit on the sum of neutrino masses is $\Sigma m_\nu < 0.064$ eV, assuming the $\Lambda$CDM model, while for DESI DR1, the upper limit is slightly higher at $\Sigma m_\nu < 0.072$ eV \cite{karim2025desi,adame2025desi}. These cosmological constraints are complementary to laboratory neutrino experiments \cite{katrin2022} and serve as crucial consistency checks for $\Lambda$CDM and its extensions.

In this work, we critically examine the status of the cosmological constant in light of DESI DR2 by analyzing several DE models beyond the $\Lambda$CDM model. We ask: Is $\Lambda$CDM truly the cornerstone of modern cosmology, or are we witnessing the first signs of a crisis in our understanding of the cosmos. In Section \ref{sec_2}, we introduce the cosmological background equations and models, providing the foundation for our analysis. Section \ref{sec_3} covers the core of this work, focusing on the datasets and methodology, particularly the use of Markov Chain Monte Carlo sampling with the publicly available \texttt{SimpleMC} cosmological inference code. Section \ref{sec_4} presents the discussion of the results of these analyzes. Finally, in Section \ref{sec_5}, we summarize our conclusions and suggest possible directions for future research.

\section{Standard Background Equations and the Cosmological Constant Model}\label{sec_2}
General Relativity (GR) satisfies the Lovelock theorem \cite{lovelock1971einstein}, which states that the only second-order field equations derivable from a scalar density in four dimensions are the Einstein equations. The Einstein Hilbert action for the concordance $\Lambda$CDM model, in units $c=\hbar=1$, is
\begin{equation}
\mathcal{S} = \frac{1}{16\pi G}\int d^4x \sqrt{-g}(R - 2\Lambda) + \mathcal{S}_m,
\end{equation}

where $\Lambda$ is the cosmological constant, $G$ is the Newtonian gravitational constant, and $\mathcal{S}_m$ represents the matter action. Variation with respect to $g_{\mu\nu}$ gives
\begin{equation}
R_{\mu\nu} - \tfrac{1}{2}R g_{\mu\nu} + \Lambda g_{\mu\nu} = 8\pi G\,T_{\mu\nu},
\end{equation}

where $T_{\mu\nu} = (\rho + p)u_\mu u_\nu + p g_{\mu\nu}$ for a perfect fluid. The Bianchi identity $\nabla^\mu G_{\mu\nu}=0$ implies $\nabla^\mu T_{\mu\nu}=0$, ensuring a constant $\Lambda$ in spacetime.

For a spatially flat Friedmann Lemaître Robertson Walker (FLRW) metric,
\begin{equation}
ds^2 = -dt^2 + a^2(t)(dr^2 + r^2 d\Omega^2),
\end{equation}

the Einstein equations reduce to
\begin{align}
3H^2 &= 8\pi G \sum_i \rho_i,\\
2\dot{H} + 3H^2 &= -8\pi G \sum_i p_i,
\end{align}

where $H=\dot{a}/a$ is the Hubble rate. Energy momentum conservation gives
\begin{equation}
\dot{\rho}_i + 3H(1+\omega_i)\rho_i = 0,
\end{equation}
whose solutions yield $\rho_m = \rho_{m0}a^{-3}$ for matter and $\rho_{\text{de}} = \rho_{\text{de},0}$ for a cosmological constant with $\omega_{\text{de}}=-1$.

The dimensionless expansion rate is then
\begin{equation}
\begin{split}
E(z) &\equiv \frac{H^2(z)}{H_0^2} = 
\Omega_m (1+z)^3 + \Omega_{\text{rad}} (1+z)^4 \\
&\quad +\, \Omega_k (1+z)^2 + \Omega_\nu f_\nu(z) 
+ \Omega_{\text{de}} f_{\text{DE}}(z).
\end{split}
\end{equation}

where $\Omega_i = \tfrac{8\pi G\rho_{i0}}{3H_0^2}$ are present-day density parameters. The functions $f_{\nu}(z) = \rho_{\nu}(z)/\rho_{\nu,0}$ and $f_{\mathrm{DE}}(z) = \rho_{\mathrm{DE}}(z)/\rho_{\mathrm{DE},0}$ describe the redshift evolution of neutrinos and DE, respectively. The neutrino density evolves as radiation at early times and as matter at late times, with the transition occurring near $(1+z) \sim m_\nu /(5 \times 10^{-4},\text{eV})$~\cite{particle2022review,lesgourgues2012neutrino}. Its present value is set by the total neutrino mass, $\Omega_\nu h^2 = \sum m_\nu / (93.14,\text{eV})$. For DE, the simplest case corresponds to a cosmological constant with $f_{\mathrm{DE}}(z)=1$. 
\subsection{Dark Energy Models with Parameterized EoS}

For a general DE component with $\omega(z) = p_{\text{DE}}/\rho_{\text{DE}}$, the energy density evolves as
\begin{equation}\label{eq_9}
f_{\text{DE}}(z) = \exp\!\left[3\int_0^z \frac{1+\omega(z')}{1+z'}\,dz'\right],
\end{equation}
which reduces to $(1+z)^{3(1+\omega)}$ for constant $\omega$ (the $\omega$CDM model). Several parameterizations of $\omega(z)$ considered in this work are listed in Table~\ref{tab_0}, including CPL, logarithmic, exponential, JBP, BA, and GEDE models. Substituting the corresponding $f_{\text{DE}}(z)$ into the Friedmann equation gives the expansion history $E(z)$ for each case.

\begin{table*}
\begin{tabular}{|c|c|c|c|}
\hline
\textbf{Parameterization} & \textbf{$\omega(z)$} & \textbf{$f_{DE}(z)$} & \textbf{Reference} \\
\hline
$\omega_0\omega_a$CDM & $\omega_0 + \frac{z}{1+z} \omega_a$ & $(1+z)^{3(1+\omega_0+\omega_a)} e^{-\frac{3\omega_a z}{1+z}}$ & \cite{chevallier2001accelerating,linder2003exploring} \\
Logarithmic & $\omega_0 + \omega_a \log(1+z)$ & $(1+z)^{3(1+\omega_0)} e^{\frac{3}{2} \omega_a(\log(1+z))^2}$ & \cite{efstathiou1999constraining,silva2012thermodynamics} \\
Exponential & $\omega_0+\omega_a\left(e^{\frac{z}{1+z}}-1\right)$ & $e^{\!\left[3 \omega_a\!\left(\frac{-z}{1+z}\right)\right]\,
(1+z)^{3(1+\omega_0+\omega_a)}\,}
e^{\!\left[3 \omega_a\!\left(\frac{1}{4(1+z)^2}+\frac{1}{2(1+z)}-\frac{3}{4}\right)\right]\,
(1+z)^{\tfrac32 \omega_a}}$ & \cite{pan2020imprints,dimakis2016general,najafi2024dynamical} \\
JBP & $\omega_0 + \frac{z}{(1+z)^2} \omega_a $ & $(1+z)^{3(1+\omega_0)} e^{\frac{3\omega_a z^2}{2(1+z)^2}}$ & \cite{jassal2005wmap} \\
BA & $\omega_0 + \frac{z(1+z)}{1+z^2} \omega_a$ & $(1+z)^{3(1+\omega_0)}(1+z^2)^{\frac{3\omega_a}{2}}$ & \cite{barboza2008parametric} \\
GEDE & $-1-\frac{\Delta}{3\ln (10)}\left[ 1+\tanh\left( \Delta \log_{10}\left(\frac{1+z}{1+z_t}\right)\right)\right]$ & $\left( \frac{1 - \tanh\left(\Delta \times \log_{10} \left( \frac{1+z}{1+z_t} \right)\right)}{1 + \tanh\left(\Delta \times \log_{10} (1 + z_t)\right)} \right)$ & \cite{li2020evidence} \\
\hline
\end{tabular}
\caption{Dark energy parameterizations with their equations of state $\omega(z)$, evolution functions $f_{DE}(z)$}\label{tab_0}
\end{table*}

\subsection{Dark Energy Models assuming a nonflat Universe}
Observations allow slight deviations from spatial flatness, with CMB data favoring a closed Universe \cite{aghanim2020planck,handley2021curvature,di2020planck} and late-time probes suggesting open Universe \cite{wu2024measuring}. To accommodate curvature effects, we consider the non-flat extensions of $\Lambda$CDM and $\omega$CDM, denoted as o$\Lambda$CDM and o$\omega$CDM, respectively. For an o$\Lambda$CDM model, the normalized Hubble parameter is given by the following:
\begin{equation}
\begin{split}
E(z) &\equiv \frac{H^2(z)}{H_0^2} = 
\Omega_m (1+z)^3 + \Omega_{rad} (1+z)^4 \\
&\quad +\, \Omega_k (1+z)^2 + \Omega_\nu f_\nu(z) 
+ \Omega_{de}
\end{split}
\end{equation}
Similarly, for the o$\omega$CDM model,
\begin{equation}
\begin{split}
E(z) &\equiv \frac{H^2(z)}{H_0^2} = 
\Omega_m (1+z)^3 + \Omega_{rad} (1+z)^4 \\
&\quad +\, \Omega_k (1+z)^2 + \Omega_\nu f_\nu(z) 
+  \Omega_{de}(1+z)^{3(1+\omega)}.
\end{split}
\end{equation}

\section{Dataset and Methodology}\label{sec_3}
We perform Bayesian parameter estimation using the \texttt{SimpleMC} cosmological inference code \cite{simplemc,aubourg2015}, which use the Metropolis Hastings Markov Chain Monte Carlo (MCMC) algorithm \cite{hastings1970monte} to explore the parameter space efficiently, with convergence tested via the Gelman–Rubin statistic $R-1$ \cite{gelman1992inference} and requiring $R-1 < 0.01$. The MCMC results are subsequently analyzed and visualized using the \texttt{GetDist} package~\cite{lewis2025getdist}. We compute Bayesian evidence ($\ln \mathcal{Z}$) using \texttt{MCEvidence} \cite{heavens2017marginal}. This quantifies the fit of the model and the model comparison is performed using the Bayes factor $B_{ab} = \mathcal{Z}_a / \mathcal{Z}_b$ or the difference in logarithmic evidence $\Delta \ln \mathcal{Z}$. A smaller $\ln \mathcal{Z}$ value indicates a statistically preferred model. We interpret the strength of evidence using the revised Jeffreys scale \cite{kass1995bayes}: $|\Delta \ln \mathcal{Z}| < 1$ (inconclusive / weak), $1 \leq |\Delta \ln \mathcal{Z}| < 3$ (moderate), $3 \leq |\Delta \ln \mathcal{Z}| < 5$ (strong), and $|\Delta \ln \mathcal{Z}| \geq 5$ (decisive). In addition to the Bayesian evidence, we also consider the difference in minimum chi-square values, defined as $\Delta\chi^2 = \chi^2_{\mathrm{Model}} - \chi^2{_{\Lambda\mathrm{CDM}}}$, to assess the relative goodness of fit between models. Our analysis uses data from Baryon Acoustic Oscillation measurements, Type Ia supernovae, and Compressed CMB likelihood:
\begin{itemize}
    \item \textbf{Baryon Acoustic Oscillation:} First we use recent baryon acoustic oscillation (BAO) measurements from over 14 million galaxies and quasars obtained by the Dark Energy Spectroscopic Instrument (DESI) Data Release 2 (DR2) \cite{karim2025desi}. Which includes various tracers such as BGS, LRGs, ELGs, QSOs, and Lyman-$\alpha$ forests. Key quantities like $D_H(z)$, $D_M(z)$, and $D_V(z)$ are used to derive ratios like $D_M/r_d$, $D_H/r_d$, and $D_V/r_d$ to constrain model parameters. The sound horizon $r_d$ is $147.09 \pm 0.2 \, \text{Mpc}$ in flat $\Lambda$CDM \cite{aghanim2020planck}.
    
    \item \textbf{Type Ia Supernovae :} Then we use three supernova compilations in our analysis. The Pantheon$^+$ sample~\cite{brout2022pantheon} contains 1701 high-quality light curves from 1550 Type Ia Supernovae (SNe Ia), with $z<0.01$ excluded, as such low redshift data are affected by significant systematic uncertainties due to peculiar velocities. The DES-SN5Y dataset~\cite{abbott2024dark} comprises 1829 photometric light curves spanning five years of the Dark Energy Survey Supernova program, including 1,635 DES-discovered events and 194 externally sourced low-$z$ supernovae from CfA and CSP. The Union3 compilation~\cite{rubin2025union} includes 2,087 SNe Ia, of which 1,363 overlap with Pantheon$^+$, offering complementary coverage. For each dataset, we marginalize over the $\mathcal{M}$ parameter to account for calibration uncertainties (see Equations A9–A12 in~\cite{goliath2001supernovae}). 
    
    \item \textbf{Compressed CMB likelihood :} Finally we use the Compressed CMB likelihood, where the CMB information is represented by the parameter vector $\mathbf{v} = \{\, \omega_b,\, \omega_{cb}, D_A(1100)/r_d\}$ in a 3×3 multivariate Gaussian likelihood~\cite{aubourg2015cosmological} known as the PLK15 likelihood in \texttt{SimpleMC}. We use a compressed CMB likelihood since dynamical DE models only affect the late-time expansion history, producing mainly geometrical effects on the CMB. The full CMB spectrum includes small non geometrical anomalies, such as the lensing amplitude and low-$\ell$ power deficit, which may reflect residual systematics and bias DE inferences. For example, Planck data alone show a $\gtrsim 2\sigma$ preference for phantom DE~\cite{escamilla2024state}, largely due to the lack of large-scale power. To avoid such biases, we rely on the cleaner geometrical information provided by the compressed CMB likelihood.
\end{itemize}

\subsection{Sum of neutrino masses}
The hot Big Bang model predicts a relic neutrino background, similar to the cosmic microwave background of photons. Neutrinos play a special role in cosmology: at early times they behave like radiation, while at late times they act as a form of matter. Because of this, they leave signatures on both the early acoustic oscillations in the primordial plasma and on the late-time growth of cosmic structures. Cosmological data are therefore sensitive to both the number of neutrino species and to the total mass of all neutrinos, \(\sum m_\nu\) (see, e.g.~\cite{collaboration2016desi}). These cosmological constraints are complementary to those obtained from laboratory experiments.

In the minimal cosmological model, we fix the total neutrino mass to \(\sum m_\nu = 0.06 \, \mathrm{eV}\). This choice assumes one massive neutrino eigenstate and two massless ones and is motivated by the lower bound from oscillation experiments. Oscillation data show that neutrinos have mass and mix between flavors, but they only measure the differences between squared masses, not the absolute mass scale. This implies that in the normal hierarchy (NH) the total mass must be at least \(\sum m_\nu \geq 0.059 \, \mathrm{eV}\), while in the inverted hierarchy (IH) the minimum is higher, \(\sum m_\nu \geq 0.10 \, \mathrm{eV}\) \cite{gonzalez2021nufit}. The actual order (NH or IH) is not yet known.  

Laboratory experiments provide independent bounds on neutrino masses. The most sensitive measurement comes from KATRIN, which studies the endpoint of the tritium \(\beta\)-decay spectrum. Its result is \(m_\beta < 0.8 \, \mathrm{eV}\) (90\% CL) \cite{bornschein2005katrin,eliasdottir2022next}, corresponding to \(\sum m_\nu \lesssim 2.4 \, \mathrm{eV}\). This limit is independent of cosmology. Together, oscillation and direct measurements imply that \(\sum m_\nu\) is between about \(0.06 \, \mathrm{eV}\) and a few eV.  

Cosmology provides a much more sensitive probe of \(\sum m_\nu\). Because most observables respond mainly to the total neutrino mass and not to the detailed mass splittings, analyses usually assume three degenerate eigenstates with equal mass. This approximation does not exactly match NH or IH, but reproduces their cosmological effects with high precision \cite{lesgourgues2006massive}. With this approach, a detection of non-zero \(\sum m_\nu\) gives the correct value in either hierarchy \cite{di2018exploring}, and if only upper limits are possible, the constraints are still valid \cite{choudhury2020updated}.

Massive neutrinos influence cosmology in two important ways. First, because they move at very high speeds, they free-stream across large distances, which prevents them from clustering on small scales. This reduces the matter power spectrum amplitude below the free-streaming length, in a nearly scale-independent way, and also produces a small shift in the BAO scale. Second, at late times neutrinos become non-relativistic (\(\sum m_\nu \gg T \simeq 10^{-3}\,\mathrm{eV}\)) and contribute to the total matter density,
\[
\Omega_\nu h^2 = \frac{\sum m_\nu}{93.14 \, \mathrm{eV}} 
\]
This changes the background expansion history and shifts the redshift of matter-\(\Lambda\) equality \cite{lesgourgues2006massive}.  

BAO data by themselves mainly measure the geometry of the Universe, and are not directly sensitive to the suppression caused by neutrino free-streaming. However, when BAO are combined with CMB data, including lensing, the constraints on \(\sum m_\nu\) become much stronger. This is because BAO help fix the late-time expansion history and break the degeneracy between \(\sum m_\nu\), the Hubble constant \(H_0\), and the matter density \(\omega_m\). Looking ahead, DESI will provide even stronger sensitivity: BAO probes the background geometry, while the full galaxy power spectrum directly captures the small-scale suppression caused by neutrino mass \cite{desi2024desi}.

Motivated by this, we extend our analysis beyond the minimal model by allowing \(\sum m_\nu\) to vary as a free parameter. We then constrain \(\sum m_\nu\) using DESI~DR2 \cite{karim2025desi} in combination with various SNe~Ia samples, including PantheonPlus, DES-SN5Y, Union3, together with the CMB dataset.
\subsection{Number of effective relativistic species}
In addition to neutrino mass, we also allow the effective number of relativistic species, $N_{\text{eff}}$, to vary. This extension captures the possible presence of additional light relics beyond the three standard-model neutrinos, such as sterile neutrinos or other forms of dark radiation. In the standard scenario $N_{\text{eff}}=3.044$ \cite{froustey2020neutrino,bennett2021towards}, but any deviation would signal new physics in the early Universe.

In this analysis, we assume a spatially flat Universe ($\Omega_{k} = 0$) for all dynamical DE models. 
When the total neutrino mass $\sum m_{\nu}$ is not varied, we set $f_{\nu} = 0$. 
The present-day radiation density parameter is defined as  $\Omega_{\mathrm{rad}} = 2.469 \times 10^{-5}\, h^{-2} \left( 1 + 0.2271\, N_{\mathrm{eff}} \right)$~\cite{komatsu2009five}, 
where $N_{\mathrm{eff}} = 3.04$ is the standard effective number of relativistic species~\cite{mangano2002precision}. 
The DE density parameter is obtained from the flatness condition,
$\Omega_{de} = 1 - \Omega_{rad} - \Omega_{m} - \Omega_{k}$, which means that both $\Omega_{rad}$ and $\Omega_{\mathrm{de}}$ are derived from the other cosmological parameters. The priors adopted for these models are summarized in Table~\ref{tab_1}.
\begin{table}
\centering
\begin{tabular}{lll}
\hline
\textbf{Model} & \textbf{Parameter} & \textbf{Prior} \\
\hline
\multirow{2}{*}{\(\Lambda\)CDM} 
& \( \Omega_{m0} \) & \( \mathcal{U}[0, 1] \) \\
& \( h = H_0/100 \) & \( \mathcal{U}[0, 1] \) \\
\hline
\multirow{2}{*}{o\(\Lambda\)CDM} 
& \( \Omega_{k0} \) & \( \mathcal{U}[-1, 1] \) \\
& \( \Omega_{m0},\ h \) & \( \mathcal{U}[0, 1] \) \\
\hline
\multirow{2}{*}{\(\omega\)CDM} 
& \( \omega_0 \) & \( \mathcal{U}[-3, 1] \) \\
& \( \Omega_{m0},\ h \) & \( \mathcal{U}[0, 1] \) \\
\hline
\multirow{3}{*}{o\(\omega\)CDM} 
& \( \omega_0 \) & \( \mathcal{U}[-3, 1] \) \\
& \( \Omega_{k0} \) & \( \mathcal{U}[-1, 1] \) \\
& \( \Omega_{m0},\ h \) & \( \mathcal{U}[0, 1] \) \\
\hline
\multirow{3}{*}{$\omega_0\omega_a$CDM} 
& \( \omega_0 \) & \( \mathcal{U}[-3, 1] \) \\
& \( \omega_a \) & \( \mathcal{U}[-3, 2] \) \\
& \( \Omega_{m0},\ h \) & \( \mathcal{U}[0, 1] \) \\
\hline
\multirow{3}{*}{Logarithmic} 
& \( \omega_0 \) & \( \mathcal{U}[-3, 1] \) \\
& \( \omega_a \) & \( \mathcal{U}[-3, 2] \) \\
& \( \Omega_{m0},\ h \) & \( \mathcal{U}[0, 1] \) \\
\hline
\multirow{3}{*}{Exponential} 
& \( \omega_0 \) & \( \mathcal{U}[-3, 1] \) \\
& \( \omega_a \) & \( \mathcal{U}[-3, 2] \) \\
& \( \Omega_{m0},\ h \) & \( \mathcal{U}[0, 1] \) \\
\hline
\multirow{3}{*}{JBP} 
& \( \omega_0 \) & \( \mathcal{U}[-3, 1] \) \\
& \( \omega_a \) & \( \mathcal{U}[-3, 2] \) \\
& \( \Omega_{m0},\ h \) & \( \mathcal{U}[0, 1] \) \\
\hline
\multirow{3}{*}{BA} 
& \( \omega_0 \) & \( \mathcal{U}[-3, 1] \) \\
& \( \omega_a \) & \( \mathcal{U}[-3, 2] \) \\
& \( \Omega_{m0},\ h \) & \( \mathcal{U}[0, 1] \) \\
\hline
\multirow{2}{*}{GDED} 
& \( \Delta \) & \( \mathcal{U}[-10, 10] \) \\
& \( \Omega_{m0},\ h \) & \( \mathcal{U}[0, 1] \)\\
\hline
\multirow{1}{*}{Neutrino Mass} 
& \( \sum m_\nu \)eV & \( > 0\) \\
\multirow{1}{*}{Relativistic Species} 
& \( N_{eff }\) & \( \mathcal{U}[3.0, 3.1] \) \\
\hline
\end{tabular}
\caption{The table shows the parameters and the priors used in our analysis for each DE model. The symbol $\mathcal{U}$ denotes that we use uniform priors, and $h \equiv H_0/100$.}\label{tab_1}
\end{table}
\begin{table*}
\setlength{\tabcolsep}{5pt}
\resizebox{\textwidth}{!}{%
\begin{tabular}{lccccccccccc}
\hline
\textbf{Dataset/Models} & $h$ & $\Omega_{m0}$ & $\Omega_k$ & $\omega_0$ & $\omega_a$ & $\Delta$ & $\Delta\chi^2$ & $\ln\mathcal{Z}$ & $|\Delta \ln \mathcal{Z}_{\Lambda\mathrm{CDM}, \mathrm{Model}}|$ & Deviation ($\sigma$) \\
\hline
\textbf{$\Lambda$CDM} \\
CMB + DESI DR2 & $0.682{\pm0.005}$ & $0.303{\pm0.006}$ & --- & --- & --- & --- & 0 & $-26.54$ & 0 & --- \\
CMB + DESI DR2 + Pantheon$^+$ & $0.680{\pm0.004}$ & $0.306{\pm0.005}$ & --- & --- & --- & --- & 0 & $-729.26$ & 0 & --- \\
CMB + DESI DR2 + DES-SN5Y & $0.677{\pm0.004}$ & $0.308{\pm0.005}$ & --- & --- & --- & --- & 0 & $-850.24$ & 0 & --- \\
CMB + DESI DR2 + Union3 & $0.680{\pm0.004}$ & $0.305{\pm0.005}$ & --- & --- & --- & --- & 0 & $-40.54$ & 0 & --- \\
\hline
\textbf{o$\Lambda$CDM} \\
CMB + DESI DR2 & $0.694{\pm0.007}$ & $0.297{\pm0.006}$ & $0.004{\pm0.002}$ & --- & --- & --- & $-2.70$ & $-29.06$ & 2.52 & --- \\
CMB + DESI DR2 + Pantheon$^+$ & $0.690{\pm0.007}$ & $0.301{\pm0.006}$ & $0.004{\pm0.002}$ & --- & --- & --- & $-2.38$ & $-731.98$ & 2.72 & --- \\
CMB + DESI DR2 + DES-SN5Y & $0.687{\pm0.006}$ & $0.303{\pm0.006}$ & $0.004{\pm0.002}$ & --- & --- & --- & $-2.07$ & $-853.37$ & 3.13 & --- \\
CMB + DESI DR2 + Union3 & $0.690{\pm0.007}$ & $0.300{\pm0.006}$ & $0.004{\pm0.002}$ & --- & --- & --- & $-2.38$ & $-43.31$ & 2.77 & --- \\
\hline
\textbf{$\omega$CDM} \\
CMB + DESI DR2 & $0.690{\pm0.011}$ & $0.299{\pm0.009}$ & --- & $-1.004{\pm0.044}$ & --- & --- & $-2.30$ & $-25.91$ & 0.62 & 0.09 \\
CMB + DESI DR2 + Pantheon$^+$ & $0.681{\pm0.007}$ & $0.306{\pm0.006}$ & --- & $-0.968{\pm0.027}$ & --- & --- & $-2.84$ & $-728.55$ & 0.71 & 1.19 \\
CMB + DESI DR2 + DES-SN5Y & $0.675{\pm0.006}$ & $0.311{\pm0.006}$ & --- & $-0.944{\pm0.026}$ & --- & --- & $-4.48$ & $-847.91$ & 2.32 & 2.15 \\
CMB + DESI DR2 + Union3 & $0.678{\pm0.008}$ & $0.308{\pm0.007}$ & --- & $-0.956{\pm0.033}$ & --- & --- & $-3.07$ & $-39.35$ & 1.19 & 1.33 \\
\hline
\textbf{o$\omega$CDM} \\
CMB + DESI DR2 & $0.692{\pm0.010}$ & $0.299{\pm0.008}$ & $0.005{\pm0.002}$ & $-0.999_{-0.039}^{+0.043}$ & --- & --- & $-2.67$ & $-31.14$ & 4.60 & 0.02 \\
CMB + DESI DR2 + Pantheon$^+$ & $0.683{\pm0.007}$ & $0.305{\pm0.006}$ & $0.005{\pm0.002}$ & $-0.964{\pm0.027}$ & --- & --- & $-3.30$ & $-733.69$ & 4.43 & 1.33 \\
CMB + DESI DR2 + DES-SN5Y & $0.678{\pm0.007}$ & $0.310{\pm0.006}$ & $0.005{\pm0.002}$ & $-0.939{\pm0.025}$ & --- & --- & $-5.03$ & $-853.04$ & 2.80 & 2.44 \\
CMB + DESI DR2 + Union3 & $0.680{\pm0.008}$ & $0.308{\pm0.007}$ & $0.005{\pm0.002}$ & $-0.950{\pm0.033}$ & --- & --- & $-3.58$ & $-44.44$ & 3.90 & 1.52 \\
\hline
\textbf{$\omega_a\omega_0$CDM} \\
CMB + DESI DR2 & $0.651{\pm0.018}$ & $0.341{\pm0.019}$ & --- & $-0.547{\pm0.190}$ & $-1.270_{-0.540}^{+0.620}$ & --- & $-3.91$ & $-24.68$ & 2.44 & 2.38 \\
CMB + DESI DR2 + Pantheon$^+$ & $0.680{\pm0.007}$ & $0.309{\pm0.007}$ & --- & $-0.876{\pm0.062}$ & $-0.410_{-0.220}^{+0.270}$ & --- & $-4.17$ & $-728.51$ & 1.46 & 2.00 \\
CMB + DESI DR2 + DES-SN5Y & $0.672{\pm0.006}$ & $0.317{\pm0.007}$ & --- & $-0.795{\pm0.063}$ & $-0.660_{-0.240}^{+0.270}$ & --- & $-7.96$ & $-845.79$ & 5.07 & 3.25 \\
CMB + DESI DR2 + Union3 & $0.664{\pm0.008}$ & $0.325{\pm0.009}$ & --- & $-0.716{\pm0.089}$ & $-0.860_{-0.290}^{+0.320}$ & --- & $-6.82$ & $-36.76$ & 4.33 & 3.19 \\
\hline
\textbf{Logarithmic} \\
CMB + DESI DR2 & $0.658{\pm0.020}$ & $0.333{\pm0.018}$ & --- & $-0.690{\pm0.150}$ & $-0.720{\pm0.480}$ & --- & $-3.52$ & $-24.84$ & 1.69 & 2.07 \\
CMB + DESI DR2 + Pantheon$^+$ & $0.679{\pm0.006}$ & $0.310{\pm0.006}$ & --- & $-0.887{\pm0.050}$ & $-0.290_{-0.140}^{+0.180}$ & --- & $-4.26$ & $-728.39$ & 0.87 & 2.26 \\
CMB + DESI DR2 + DES-SN5Y & $0.673{\pm0.006}$ & $0.317{\pm0.007}$ & --- & $-0.822{\pm0.055}$ & $-0.430_{-0.160}^{+0.190}$ & --- & $-7.83$ & $-845.75$ & 4.48 & 3.24 \\
CMB + DESI DR2 + Union3 & $0.666{\pm0.008}$ & $0.324{\pm0.009}$ & --- & $-0.760{\pm0.075}$ & $-0.570_{-0.190}^{+0.230}$ & --- & $-6.64$ & $-36.80$ & 3.73 & 3.20 \\
\hline
\textbf{Exponential} \\
CMB + DESI DR2 & $0.626_{-0.021}^{+0.023}$ & $0.369_{-0.023}^{+0.031}$ & --- & $-0.755_{-0.11}^{+0.089}$ & $-0.950_{-0.30}^{+0.41}$ & --- & $-4.32$ & $-24.10$ & 2.36 & 2.46 \\
CMB + DESI DR2 + Pantheon$^+$ & $0.675{\pm0.006}$ & $0.312{\pm0.006}$ & --- & $-0.972{\pm0.026}$ & $-0.221_{-0.099}^{+0.11}$ & --- & $-3.38$ & $-727.80$ & 0.51 & 1.08 \\
CMB + DESI DR2 + DES-SN5Y & $0.668{\pm0.006}$ & $0.320{\pm0.006}$ & --- & $-0.942{\pm0.026}$ & $-0.350_{-0.10}^{+0.12}$ & --- & $-7.65$ & $-845.17$ & 3.81 & 2.23 \\
CMB + DESI DR2 + Union3 & $0.657{\pm0.008}$ & $0.330{\pm0.009}$ & --- & $-0.899{\pm0.038}$ & $-0.480{\pm0.15}$ & --- & $-6.79$ & $-36.21$ & 3.55 & 2.66 \\
\hline
\textbf{JBP} \\
CMB + DESI DR2 & $0.662_{-0.019}^{+0.012}$ & $0.326_{-0.014}^{+0.019}$ & --- & $-0.630_{-0.110}^{+0.230}$ & $-1.77_{-1.10}^{+0.46}$ & --- & $-3.01$ & $-24.70$ & 2.73 & 2.18 \\
CMB + DESI DR2 + Pantheon$^+$ & $0.679{\pm0.006}$ & $0.309{\pm0.006}$ & --- & $-0.864{\pm0.080}$ & $-0.670{\pm0.490}$ & --- & $-3.71$ & $-728.08$ & 1.43 & 1.70 \\
CMB + DESI DR2 + DES-SN5Y & $0.671{\pm0.006}$ & $0.317{\pm0.007}$ & --- & $-0.727{\pm0.087}$ & $-1.370{\pm0.540}$ & --- & $-7.96$ & $-845.24$ & 5.91 & 3.14 \\
CMB + DESI DR2 + Union3 & $0.662{\pm0.009}$ & $0.325{\pm0.009}$ & --- & $-0.621_{-0.093}^{+0.130}$ & $-1.860_{-0.730}^{+0.490}$ & --- & $-6.79$ & $-36.64$ & 5.30 & 3.40 \\
\hline
\textbf{BA} \\
CMB + DESI DR2 & $0.648{\pm0.020}$ & $0.344_{-0.024}^{+0.021}$ & --- & $-0.600_{-0.200}^{+0.180}$ & $-0.650_{-0.270}^{+0.330}$ & --- & $-3.89$ & $-23.81$ & 1.86 & 2.11 \\
CMB + DESI DR2 + Pantheon$^+$ & $0.679{\pm0.006}$ & $0.310{\pm0.006}$ & --- & $-0.889{\pm0.051}$ & $-0.210_{-0.110}^{+0.120}$ & --- & $-4.25$ & $-727.83$ & 0.75 & 2.18 \\
CMB + DESI DR2 + DES-SN5Y & $0.672{\pm0.006}$ & $0.317{\pm0.007}$ & --- & $-0.822{\pm0.050}$ & $-0.320_{-0.120}^{+0.140}$ & --- & $-7.92$ & $-844.33$ & 4.45 & 3.56 \\
CMB + DESI DR2 + Union3 & $0.664{\pm0.008}$ & $0.325{\pm0.009}$ & --- & $-0.752{\pm0.078}$ & $-0.420_{-0.150}^{+0.160}$ & --- & $-6.87$ & $-35.24$ & 3.78 & 3.18 \\
\hline
\textbf{GEDE} \\
CMB + DESI DR2 & $0.668{\pm0.010}$ & $0.300{\pm0.008}$ & --- & --- & --- & $0.099_{-0.320}^{+0.280}$ & $1.10$ & $-25.79$ & 0.74 & 0.33 \\
CMB + DESI DR2 + Pantheon$^+$ & $0.678{\pm0.006}$ & $0.307{\pm0.006}$ & --- & --- & --- & $-0.160{\pm0.19}$ & $1.16$ & $-728.56$ & 0.69 & 0.84 \\
CMB + DESI DR2 + DES-SN5Y & $0.671{\pm0.006}$ & $0.313{\pm0.006}$ & --- & --- & --- & $-0.340_{-0.160}^{+0.140}$ & $-0.24$ & $-848.03$ & 2.20 & 2.27 \\
CMB + DESI DR2 + Union3 & $0.673{\pm0.008}$ & $0.311{\pm0.007}$ & --- & --- & --- & $-0.310{\pm0.21}$ & $-1.04$ & $-39.42$ & 1.12 & 1.48 \\
\hline
\end{tabular}
}
\caption{This table presents the numerical values obtained for the o$\Lambda$CDM ,$\omega$CDM , o$\omega$CDM, $\omega_a\omega_0$CDM, Logarithmic, Exponential, JBP, BA, and GEDE models at the 68\% (1$\sigma$) confidence level, using different combinations of DESI DR2 BAO datasets with the CMB and various SNe~Ia samples.}
\label{tab_2}
\end{table*}


\begin{table*}
\setlength{\tabcolsep}{5pt}
\resizebox{\textwidth}{!}{%
\begin{tabular}{lccccccccccc}
\hline
\textbf{Dataset/Models} & $h$ & $\Omega_{m}$ & $\Omega_{k}$ & $\omega_0$ & $\omega_a$ & $\Delta$ & $\sum m_\nu$ [eV] & $\Delta\chi^2$ & $\ln\mathcal{Z}$ & $|\Delta\ln\mathcal{Z}_{\Lambda\mathrm{CDM},\mathrm{Model}}|$ & Deviation ($\sigma$) \\
\hline
\textbf{$\Lambda$CDM + $\sum m_\nu$} \\
CMB + DESI DR2 & $0.682{\pm0.005}$ & $0.303{\pm0.006}$ & --- & --- & --- & --- & $<0.066$ & 0 & $-28.37$ & 0 & --- \\
CMB + DESI DR2 + Pantheon$^+$ & $0.679{\pm0.005}$ & $0.306{\pm0.005}$ & --- & --- & --- & --- & $<0.073$ & 0 & $-731.07$ & 0 & --- \\
CMB + DESI DR2 + DES-SN5Y & $0.677{\pm0.005}$ & $0.309{\pm0.006}$ & --- & --- & --- & --- & $<0.086$ & 0 & $-852.11$ & 0 & --- \\
CMB + DESI DR2 + Union3 & $0.679{\pm0.005}$ & $0.306{\pm0.006}$ & --- & --- & --- & --- & $<0.074$ & 0 & $-42.34$ & 0 & --- \\
\hline
\textbf{o$\Lambda$CDM + $\sum m_\nu$} \\
CMB + DESI DR2 & $0.693{\pm0.007}$ & $0.300{\pm0.007}$ & $0.0077^{+0.0028}_{-0.0040}$ & --- & --- & --- & $<0.263$ & $-1.79$ & $-30.04$ & 1.67 & --- \\
CMB + DESI DR2 + Pantheon$^+$ & $0.690{\pm0.006}$ & $0.305{\pm0.006}$ & $0.0084^{+0.0031}_{-0.0043}$ & --- & --- & --- & $<0.267$ & $-1.75$ & $-732.67$ & 1.60 & --- \\
CMB + DESI DR2 + DES-SN5Y & $0.687{\pm0.006}$ & $0.309{\pm0.007}$ & $0.0095^{+0.0037}_{-0.0045}$ & --- & --- & --- & $<0.340$ & $-2.06$ & $-853.50$ & 1.39 & --- \\
CMB + DESI DR2 + Union3 & $0.690{\pm0.007}$ & $0.304{\pm0.007}$ & $0.0083^{+0.0032}_{-0.0041}$ & --- & --- & --- & $<0.260$ & $-1.77$ & $-43.99$ & 1.65 & --- \\
\hline
\textbf{$\omega$CDM + $\sum m_\nu$} \\
CMB + DESI DR2 & $0.687{\pm0.010}$ & $0.300{\pm0.008}$ & --- & $-1.022{\pm0.042}$ & --- & --- & $<0.075$ & 0.03 & $-30.42$ & 2.05 & 0.52 \\
CMB + DESI DR2 + Pantheon$^+$ & $0.676{\pm0.006}$ & $0.308{\pm0.006}$ & --- & $-0.979{\pm0.026}$ & --- & --- & $<0.062$ & $-0.65$ & $-733.35$ & 2.28 & 0.81 \\
CMB + DESI DR2 + DES-SN5Y & $0.670{\pm0.006}$ & $0.313{\pm0.006}$ & --- & $-0.953{\pm0.024}$ & --- & --- & $<0.057$ & $-2.38$ & $-852.95$ & 0.84 & 1.96 \\
CMB + DESI DR2 + Union3 & $0.674{\pm0.007}$ & $0.309{\pm0.007}$ & --- & $-0.969{\pm0.030}$ & --- & --- & $<0.056$ & $-0.86$ & $-44.40$ & 2.06 & 1.03 \\
\hline
\textbf{o$\omega$CDM + $\sum m_\nu$} \\
CMB + DESI DR2 & $0.700^{+0.012}_{-0.014}$ & $0.299{\pm0.009}$ & $0.0103^{+0.0040}_{-0.0054}$ & $-1.057^{+0.076}_{-0.054}$ & --- & --- & $<0.520$ & $-1.77$ & $-31.47$ & 3.10 & 0.81 \\
CMB + DESI DR2 + Pantheon$^+$ & $0.685{\pm0.007}$ & $0.308{\pm0.007}$ & $0.0081^{+0.0028}_{-0.0042}$ & $-0.979^{+0.031}_{-0.028}$ & --- & --- & $<0.260$ & $-2.68$ & $-734.76$ & 3.69 & 0.71 \\
CMB + DESI DR2 + DES-SN5Y & $0.678{\pm0.007}$ & $0.313^{+0.0067}_{-0.0076}$ & $0.0079^{+0.0025}_{-0.0043}$ & $-0.949^{+0.032}_{-0.027}$ & --- & --- & $<0.197$ & $-4.66$ & $-854.33$ & 2.22 & 1.73 \\
CMB + DESI DR2 + Union3 & $0.683{\pm0.008}$ & $0.309{\pm0.007}$ & $0.0081^{+0.0029}_{-0.0042}$ & $-0.969^{+0.040}_{-0.034}$ & --- & --- & $<0.257$ & $-2.99$ & $-45.72$ & 3.38 & 0.84 \\
\hline
\textbf{$\omega_0\omega_a$CDM + $\sum m_\nu$} \\
CMB + DESI DR2 & $0.629{\pm0.017}$ & $0.364{\pm0.021}$ & --- & $-0.380{\pm0.180}$ & $-1.91{\pm0.55}$ & --- & $<0.127$ & $-3.55$ & $-24.58$ & 3.79 & 3.44 \\
CMB + DESI DR2 + Pantheon$^+$ & $0.673{\pm0.006}$ & $0.313{\pm0.007}$ & --- & $-0.862{\pm0.056}$ & $-0.53_{-0.22}^{+0.25}$ & --- & $<0.101$ & $-2.67$ & $-731.24$ & 0.17 & 2.46 \\
CMB + DESI DR2 + DES-SN5Y & $0.666{\pm0.006}$ & $0.320{\pm0.006}$ & --- & $-0.779{\pm0.057}$ & $-0.79_{-0.22}^{+0.27}$ & --- & $<0.123$ & $-7.01$ & $-847.73$ & 4.38 & 3.88 \\
CMB + DESI DR2 + Union3 & $0.656{\pm0.008}$ & $0.337{\pm0.009}$ & --- & $-0.674{\pm0.090}$ & $-1.09{\pm0.33}$ & --- & $<0.126$ & $-5.94$ & $-38.20$ & 4.14 & 3.62 \\
\hline
\textbf{Logarithmic + $\sum m_\nu$} \\
CMB + DESI DR2 & $0.625{\pm0.026}$ & $0.371^{+0.030}_{-0.036}$ & --- & $-0.37^{+0.26}_{-0.30}$ & $-1.60_{-0.67}^{+0.76}$ & --- & $<0.123$ & $-3.59$ & $-24.70$ & 3.67 & 2.25 \\
CMB + DESI DR2 + Pantheon$^+$ & $0.675{\pm0.007}$ & $0.313{\pm0.007}$ & --- & $-0.870{\pm0.052}$ & $-0.40^{+0.19}_{-0.16}$ & --- & $<0.067$ & $-3.42$ & $-730.81$ & 0.26 & 2.50 \\
CMB + DESI DR2 + DES-SN5Y & $0.668{\pm0.006}$ & $0.320{\pm0.007}$ & --- & $-0.803{\pm0.055}$ & $-0.56_{-0.18}^{+0.22}$ & --- & $<0.072$ & $-7.34$ & $-847.76$ & 4.35 & 3.58 \\
CMB + DESI DR2 + Union3 & $0.657{\pm0.009}$ & $0.331{\pm0.011}$ & --- & $-0.707{\pm0.086}$ & $-0.79_{-0.24}^{+0.29}$ & --- & $<0.085$ & $-6.17$ & $-38.31$ & 4.03 & 3.41 \\
\hline
\textbf{Exponential + $\sum m_\nu$} \\
CMB + DESI DR2 & $0.617{\pm0.029}$ & $0.380^{+0.035}_{-0.042}$ & --- & $-0.74^{+0.12}_{-0.13}$ & $-1.13^{+0.57}_{-0.46}$ & --- & $<0.112$ & $-3.63$ & $-24.58$ & 3.79 & 2.08 \\
CMB + DESI DR2 + Pantheon$^+$ & $0.675{\pm0.007}$ & $0.312{\pm0.007}$ & --- & $-0.974{\pm0.030}$ & $-0.23^{+0.13}_{-0.11}$ & --- & $<0.074$ & $-3.30$ & $-731.26$ & 0.19 & 0.87 \\
CMB + DESI DR2 + DES-SN5Y & $0.667{\pm0.006}$ & $0.320{\pm0.007}$ & --- & $-0.946{\pm0.030}$ & $-0.36^{+0.14}_{-0.12}$ & --- & $<0.067$ & $-7.38$ & $-848.10$ & 4.01 & 1.80 \\
CMB + DESI DR2 + Union3 & $0.657{\pm0.009}$ & $0.331{\pm0.011}$ & --- & $-0.906{\pm0.040}$ & $-0.50^{+0.18}_{-0.16}$ & --- & $<0.077$ & $-6.16$ & $-38.62$ & 3.72 & 2.35 \\
\hline
\textbf{JBP + $\sum m_\nu$} \\
CMB + DESI DR2 & $0.660^{+0.012}_{-0.019}$ & $0.327^{+0.019}_{-0.013}$ & --- & $-0.62^{+0.23}_{-0.096}$ & $-1.86^{+0.35}_{-1.10}$ & --- & $<0.054$ & $-2.82$ & $-26.70$ & 1.67 & 2.33 \\
CMB + DESI DR2 + Pantheon$^+$ & $0.676{\pm0.007}$ & $0.310{\pm0.007}$ & --- & $-0.853{\pm0.081}$ & $-0.80{\pm0.52}$ & --- & $<0.055$ & $-2.77$ & $-730.65$ & 0.42 & 1.81 \\
CMB + DESI DR2 + DES-SN5Y & $0.667{\pm0.006}$ & $0.319{\pm0.007}$ & --- & $-0.712{\pm0.088}$ & $-1.54{\pm0.55}$ & --- & $<0.059$ & $-7.25$ & $-847.09$ & 5.02 & 3.27 \\
CMB + DESI DR2 + Union3 & $0.659{\pm0.009}$ & $0.326{\pm0.009}$ & --- & $-0.613^{+0.13}_{-0.087}$ & $-1.97^{+0.44}_{-0.72}$ & --- & $<0.056$ & $-5.91$ & $-38.10$ & 4.24 & 3.57 \\
\hline
\textbf{BA + $\sum m_\nu$} \\
CMB + DESI DR2 & $0.625^{+0.025}_{-0.022}$ & $0.369^{+0.024}_{-0.034}$ & --- & $-0.40^{+0.20}_{-0.28}$ & $-1.06^{+0.51}_{-0.34}$ & --- & $<0.109$ & $-3.62$ & $-24.87$ & 3.50 & 2.50 \\
CMB + DESI DR2 + Pantheon$^+$ & $0.675{\pm0.007}$ & $0.312{\pm0.007}$ & --- & $-0.878{\pm0.053}$ & $-0.27^{+0.15}_{-0.12}$ & --- & $<0.062$ & $-3.39$ & $-731.01$ & 0.06 & 2.18 \\
CMB + DESI DR2 + DES-SN5Y & $0.667{\pm0.006}$ & $0.320{\pm0.007}$ & --- & $-0.806{\pm0.054}$ & $-0.40^{+0.15}_{-0.13}$ & --- & $<0.072$ & $-7.32$ & $-848.13$ & 3.98 & 1.29 \\
CMB + DESI DR2 + Union3 & $0.658{\pm0.009}$ & $0.330{\pm0.009}$ & --- & $-0.718{\pm0.079}$ & $-0.54^{+0.18}_{-0.16}$ & --- & $<0.078$ & $-6.21$ & $-38.77$ & 3.57 & 3.57 \\
\hline
\textbf{GEDE + $\sum m_\nu$} \\
CMB + DESI DR2 & $0.691{\pm0.008}$ & $0.296{\pm0.007}$ & --- & --- & --- & $0.33{\pm0.24}$ & $<0.043$ & 0.09 & $-29.35$ & 0.98 & 1.38 \\
CMB + DESI DR2 + Pantheon$^+$ & $0.676{\pm0.006}$ & $0.307{\pm0.006}$ & --- & --- & --- & $-0.14^{+0.17}_{-0.19}$ & $<0.030$ & $-0.58$ & $-732.21$ & 1.14 & 0.78 \\
CMB + DESI DR2 + DES-SN5Y & $0.669{\pm0.006}$ & $0.315{\pm0.006}$ & --- & --- & --- & $-0.33{\pm0.16}$ & $<0.029$ & $-2.51$ & $-851.66$ & 0.45 & 2.06 \\
CMB + DESI DR2 + Union3 & $0.673{\pm0.007}$ & $0.310{\pm0.007}$ & --- & --- & --- & $-0.23{\pm0.22}$ & $<0.031$ & $-0.86$ & $-43.07$ & 0.73 & 1.05 \\
\hline
\end{tabular}
}
\caption{This table presents the numerical values obtained for the o$\Lambda$CDM + $\sum m_\nu$, $\omega$CDM + $\sum m_\nu$, o$\omega$CDM + $\sum m_\nu$, $\omega_a\omega_0$CDM + $\sum m_\nu$, Logarithmic + $\sum m_\nu$, Exponential + $\sum m_\nu$, JBP + $\sum m_\nu$, BA + $\sum m_\nu$, and GEDE + $\sum m_\nu$ models at the 68\% (1$\sigma$) confidence level, using different combinations of DESI DR2 BAO datasets with the CMB and various SNe~Ia samples.}
\label{tab_3}
\end{table*}

\begin{table*}
\setlength{\tabcolsep}{5pt}
\resizebox{\textwidth}{!}{%
\begin{tabular}{lcccccccccccc}
\hline
\textbf{Dataset/Models} & $h$ & $\Omega_m$ & $\Omega_k$ & $\omega_0$ & $\omega_a$ & $\Delta$ & $N_{\text{eff}}$ & $\Delta\chi^2$ & $\ln\mathcal{Z}$ & $|\Delta\ln\mathcal{Z}_{\Lambda\mathrm{CDM},\mathrm{Model}}|$ & Deviation ($\sigma$) \\
\hline
\textbf{$\Lambda$CDM + $N_{\text{eff}}$} \\
CMB + DESI DR2 & $0.684{\pm0.005}$ & $0.302{\pm0.006}$ & --- & --- & --- & --- & $3.060{\pm0.02}$ & 0 & $-27.90$ & 0 & --- \\
CMB + DESI DR2 + Pantheon$^+$ & $0.681{\pm0.005}$ & $0.305{\pm0.005}$ & --- & --- & --- & --- & $3.046{\pm0.02}$ & 0 & $-730.58$ & 0 & --- \\
CMB + DESI DR2 + DES-SN5Y & $0.679{\pm0.004}$ & $0.307{\pm0.005}$ & --- & --- & --- & --- & $3.057{\pm0.02}$ & 0 & $-851.69$ & 0 & --- \\
CMB + DESI DR2 + Union3 & $0.681{\pm0.005}$ & $0.305{\pm0.006}$ & --- & --- & --- & --- & $3.059{\pm0.02}$ & 0 & $-41.85$ & 0 & --- \\
\hline
\textbf{o$\Lambda$CDM + $N_{\text{eff}}$} \\
CMB + DESI DR2 & $0.694{\pm0.007}$ & $0.297{\pm0.006}$ & $0.004{\pm0.002}$ & --- & --- & --- & $3.050{\pm0.02}$ & $-1.87$ & $-31.23$ & 3.33 & --- \\
CMB + DESI DR2 + Pantheon$^+$ & $0.690{\pm0.006}$ & $0.301{\pm0.006}$ & $0.004{\pm0.002}$ & --- & --- & --- & $3.048{\pm0.02}$ & $-1.75$ & $-734.20$ & 3.62 & --- \\
CMB + DESI DR2 + DES-SN5Y & $0.687{\pm0.006}$ & $0.304{\pm0.006}$ & $0.004{\pm0.002}$ & --- & --- & --- & $3.046{\pm0.02}$ & $-1.69$ & $-855.53$ & 3.84 & --- \\
CMB + DESI DR2 + Union3 & $0.690{\pm0.006}$ & $0.300{\pm0.006}$ & $0.004{\pm0.002}$ & --- & --- & --- & $3.048{\pm0.02}$ & $-1.74$ & $-45.51$ & 3.66 & --- \\
\hline
\textbf{$\omega$CDM + $N_{\text{eff}}$} \\
CMB + DESI DR2 & $0.687{\pm0.010}$ & $0.300{\pm0.009}$ & --- & $-1.01{\pm0.04}$ & --- & --- & $3.060{\pm0.02}$ & 0.01 & $-30.77$ & 2.87 & 2.50 \\
CMB + DESI DR2 + Pantheon$^+$ & $0.676{\pm0.006}$ & $0.308{\pm0.006}$ & --- & $-0.977{\pm0.025}$ & --- & --- & $3.064{\pm0.02}$ & $-0.60$ & $-733.59$ & 3.01 & 0.92 \\
CMB + DESI DR2 + DES-SN5Y & $0.670{\pm0.006}$ & $0.313{\pm0.006}$ & --- & $-0.953{\pm0.025}$ & --- & --- & $3.065{\pm0.02}$ & $-2.29$ & $-853.15$ & 1.46 & 1.88 \\
CMB + DESI DR2 + Union3 & $0.674{\pm0.007}$ & $0.310{\pm0.007}$ & --- & $-0.969{\pm0.031}$ & --- & --- & $3.064{\pm0.02}$ & $-0.78$ & $-44.52$ & 2.67 & 1.00 \\
\hline
\textbf{o$\omega$CDM + $N_{\text{eff}}$} \\
CMB + DESI DR2 & $0.692{\pm0.011}$ & $0.299{\pm0.008}$ & $0.004{\pm0.002}$ & $-0.997{\pm0.042}$ & --- & --- & $3.050{\pm0.02}$ & $-1.86$ & $-33.40$ & 5.50 & 0.92 \\
CMB + DESI DR2 + Pantheon$^+$ & $0.684{\pm0.007}$ & $0.305{\pm0.006}$ & $0.005{\pm0.002}$ & $-0.965{\pm0.027}$ & --- & --- & $3.051{\pm0.02}$ & $-2.73$ & $-735.87$ & 5.29 & 1.30 \\
CMB + DESI DR2 + DES-SN5Y & $0.678{\pm0.007}$ & $0.310{\pm0.006}$ & $0.005{\pm0.002}$ & $-0.939{\pm0.025}$ & --- & --- & $3.053{\pm0.02}$ & $-4.64$ & $-855.19$ & 3.50 & 2.44 \\
CMB + DESI DR2 + Union3 & $0.681{\pm0.008}$ & $0.308{\pm0.007}$ & $0.005{\pm0.002}$ & $-0.951{\pm0.032}$ & --- & --- & $3.052{\pm0.02}$ & $-3.08$ & $-44.00$ & 2.15 & 1.53 \\
\hline
\textbf{$\omega_0\omega_a$CDM + $N_{\text{eff}}$} \\
CMB + DESI DR2 & $0.638^{+0.019}_{-0.026}$ & $0.354^{+0.029}_{-0.024}$ & --- & $-0.47^{+0.28}_{-0.24}$ & $-1.61^{+0.73}_{-0.82}$ & --- & $3.051{\pm0.02}$ & $-3.65$ & $-25.54$ & 2.36 & 2.04 \\
CMB + DESI DR2 + Pantheon$^+$ & $0.674{\pm0.006}$ & $0.313{\pm0.006}$ & --- & $-0.862{\pm0.055}$ & $-0.50{\pm0.22}$ & --- & $3.049{\pm0.02}$ & $-2.70$ & $-731.57$ & 0.99 & 2.51 \\
CMB + DESI DR2 + DES-SN5Y & $0.667{\pm0.006}$ & $0.320{\pm0.006}$ & --- & $-0.779{\pm0.062}$ & $-0.76{\pm0.25}$ & --- & $3.054{\pm0.02}$ & $-7.04$ & $-848.26$ & 3.43 & 3.56 \\
CMB + DESI DR2 + Union3 & $0.658{\pm0.009}$ & $0.330{\pm0.010}$ & --- & $-0.688{\pm0.095}$ & $-1.01^{+0.35}_{-0.31}$ & --- & $3.053{\pm0.02}$ & $-5.98$ & $-38.70$ & 3.15 & 3.28 \\
\hline
\textbf{Logarithmic + $N_{\text{eff}}$} \\
CMB + DESI DR2 & $0.645^{+0.023}_{-0.019}$ & $0.346^{+0.020}_{-0.028}$ & --- & $-0.57^{+0.17}_{-0.25}$ & $-1.03^{+0.59}_{-0.37}$ & --- & $3.045{\pm0.02}$ & $-3.43$ & $-25.87$ & 2.03 & 2.05 \\
CMB + DESI DR2 + Pantheon$^+$ & $0.676{\pm0.006}$ & $0.312{\pm0.006}$ & --- & $-0.874{\pm0.052}$ & $-0.38^{+0.18}_{-0.15}$ & --- & $3.053{\pm0.02}$ & $-3.31$ & $-730.96$ & 0.38 & 2.42 \\
CMB + DESI DR2 + DES-SN5Y & $0.668{\pm0.006}$ & $0.319{\pm0.006}$ & --- & $-0.804{\pm0.053}$ & $-0.54^{+0.20}_{-0.16}$ & --- & $3.051{\pm0.02}$ & $-7.36$ & $-847.98$ & 3.71 & 3.70 \\
CMB + DESI DR2 + Union3 & $0.660{\pm0.008}$ & $0.328{\pm0.009}$ & --- & $-0.727{\pm0.080}$ & $-0.71^{+0.25}_{-0.21}$ & --- & $3.049{\pm0.02}$ & $-6.20$ & $-38.58$ & 3.27 & 3.41 \\
\hline
\textbf{Exponential + $N_{\text{eff}}$} \\
CMB + DESI DR2 & $0.631{\pm0.023}$ & $0.362^{+0.026}_{-0.030}$ & --- & $-0.78{\pm0.10}$ & $-0.87^{+0.40}_{-0.33}$ & --- & $3.046{\pm0.02}$ & $-3.57$ & $-26.05$ & 1.85 & 2.20 \\
CMB + DESI DR2 + Pantheon$^+$ & $0.675{\pm0.006}$ & $0.312{\pm0.006}$ & --- & $-0.971{\pm0.028}$ & $-0.23{\pm0.10}$ & --- & $3.055{\pm0.02}$ & $-2.96$ & $-731.85$ & 1.27 & 1.04 \\
CMB + DESI DR2 + DES-SN5Y & $0.668{\pm0.006}$ & $0.319{\pm0.007}$ & --- & $-0.942{\pm0.026}$ & $-0.34^{+0.12}_{-0.11}$ & --- & $3.053{\pm0.02}$ & $-7.33$ & $-848.55$ & 3.14 & 2.23 \\
CMB + DESI DR2 + Union3 & $0.658{\pm0.009}$ & $0.330{\pm0.010}$ & --- & $-0.902{\pm0.037}$ & $-0.47^{+0.16}_{-0.14}$ & --- & $3.052{\pm0.02}$ & $-6.19$ & $-39.11$ & 2.74 & 2.65 \\
\hline
\textbf{JBP + $N_{\text{eff}}$} \\
CMB + DESI DR2 & $0.664^{+0.013}_{-0.021}$ & $0.323^{+0.020}_{-0.015}$ & --- & $-0.68^{+0.27}_{-0.12}$ & $-1.63^{+0.53}_{-1.30}$ & --- & $3.059{\pm0.02}$ & $-2.51$ & $-26.82$ & 1.08 & 1.64 \\
CMB + DESI DR2 + Pantheon$^+$ & $0.676{\pm0.006}$ & $0.310{\pm0.006}$ & --- & $-0.851{\pm0.082}$ & $-0.83{\pm0.51}$ & --- & $3.058{\pm0.02}$ & $-2.23$ & $-730.92$ & 0.34 & 1.82 \\
CMB + DESI DR2 + DES-SN5Y & $0.667{\pm0.006}$ & $0.319{\pm0.006}$ & --- & $-0.703{\pm0.086}$ & $-1.60{\pm0.53}$ & --- & $3.056{\pm0.02}$ & $-6.82$ & $-847.48$ & 4.21 & 3.45 \\
CMB + DESI DR2 + Union3 & $0.659{\pm0.009}$ & $0.327{\pm0.009}$ & --- & $-0.614^{+0.13}_{-0.09}$ & $-1.98^{+0.45}_{-0.76}$ & --- & $3.057{\pm0.02}$ & $-5.62$ & $-38.88$ & 2.97 & 3.48 \\
\hline
\textbf{BA + $N_{\text{eff}}$} \\
CMB + DESI DR2 & $0.632{\pm0.020}$ & $0.361^{+0.022}_{-0.025}$ & --- & $-0.46^{+0.18}_{-0.21}$ & $-0.91^{+0.36}_{-0.29}$ & --- & $3.046{\pm0.02}$ & $-3.65$ & $-25.80$ & 2.10 & 2.77 \\
CMB + DESI DR2 + Pantheon$^+$ & $0.675{\pm0.006}$ & $0.312{\pm0.007}$ & --- & $-0.876{\pm0.052}$ & $-0.27^{+0.13}_{-0.11}$ & --- & $3.054{\pm0.02}$ & $-3.19$ & $-731.24$ & 0.66 & 2.38 \\
CMB + DESI DR2 + DES-SN5Y & $0.668{\pm0.006}$ & $0.319{\pm0.007}$ & --- & $-0.810{\pm0.057}$ & $-0.38^{+0.15}_{-0.13}$ & --- & $3.053{\pm0.02}$ & $-7.29$ & $-848.07$ & 3.62 & 3.33 \\
CMB + DESI DR2 + Union3 & $0.659{\pm0.009}$ & $0.330{\pm0.010}$ & --- & $-0.722{\pm0.080}$ & $-0.51{\pm0.16}$ & --- & $3.051{\pm0.02}$ & $-6.18$ & $-38.88$ & 2.97 & 3.48 \\
\hline
\textbf{GEDE + $N_{\text{eff}}$} \\
CMB + DESI DR2 & $0.687{\pm0.010}$ & $0.299{\pm0.008}$ & --- & --- & --- & $0.26{\pm0.29}$ & $3.063{\pm0.02}$ & 1.04 & $-29.85$ & 1.95 & 0.90 \\
CMB + DESI DR2 + Pantheon$^+$ & $0.675{\pm0.006}$ & $0.308{\pm0.006}$ & --- & --- & --- & $-0.09{\pm0.18}$ & $3.067{\pm0.02}$ & 0.76 & $-733.09$ & 2.51 & 0.50 \\
CMB + DESI DR2 + DES-SN5Y & $0.668{\pm0.006}$ & $0.314{\pm0.006}$ & --- & --- & --- & $-0.28{\pm0.17}$ & $3.068{\pm0.02}$ & $-0.95$ & $-852.69$ & 1.00 & 1.65 \\
CMB + DESI DR2 + Union3 & $0.673{\pm0.008}$ & $0.310{\pm0.007}$ & --- & --- & --- & $-0.14{\pm0.23}$ & $3.066{\pm0.02}$ & $-0.55$ & $-44.00$ & 2.15 & 0.61 \\
\hline
\end{tabular}
}
\caption{This table presents the numerical values obtained for the o$\Lambda$CDM + $N_{\text{eff}}$, $\omega$CDM + $N_{\text{eff}}$, o$\omega$CDM + $N_{\text{eff}}$, $\omega_a\omega_0$CDM + $N_{\text{eff}}$, Logarithmic + $N_{\text{eff}}$, Exponential + $N_{\text{eff}}$, JBP + $N_{\text{eff}}$, BA + $N_{\text{eff}}$, and GEDE + $N_{\text{eff}}$ models at the 68\% (1$\sigma$) confidence level, using different combinations of DESI DR2 BAO datasets with the CMB and various SNe~Ia samples.}
\label{tab_4}
\end{table*}
\begin{figure*}
\centering
\includegraphics[scale=0.60]{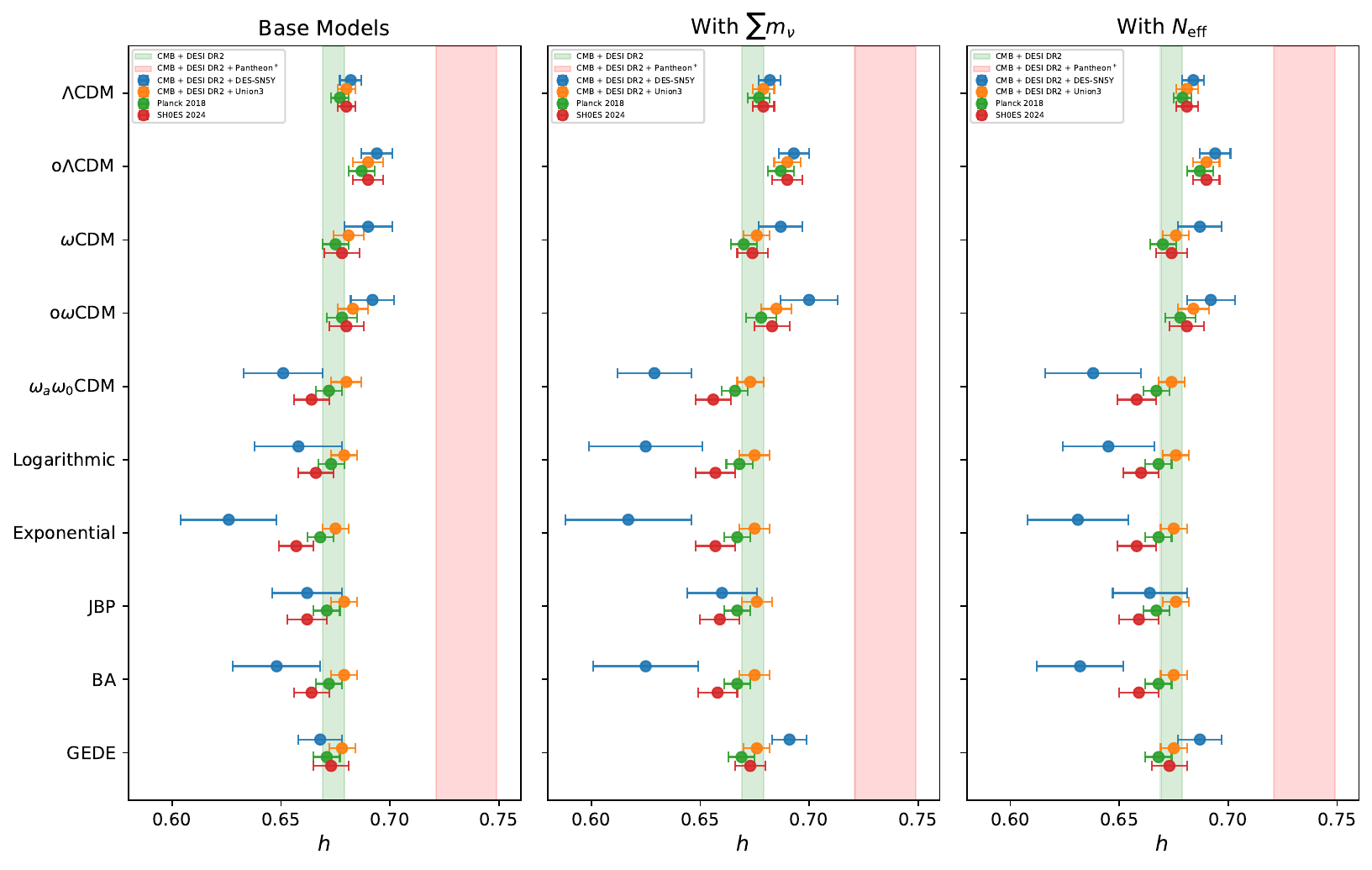}
\caption{The figure shows the summary of inferred values of $h$ for different cosmological models, with and without varying $\sum m_\nu$ and $N_\mathrm{eff}$}\label{fig_h}
\end{figure*}

\begin{figure*}
\begin{subfigure}{.3\textwidth}
\includegraphics[width=\linewidth]{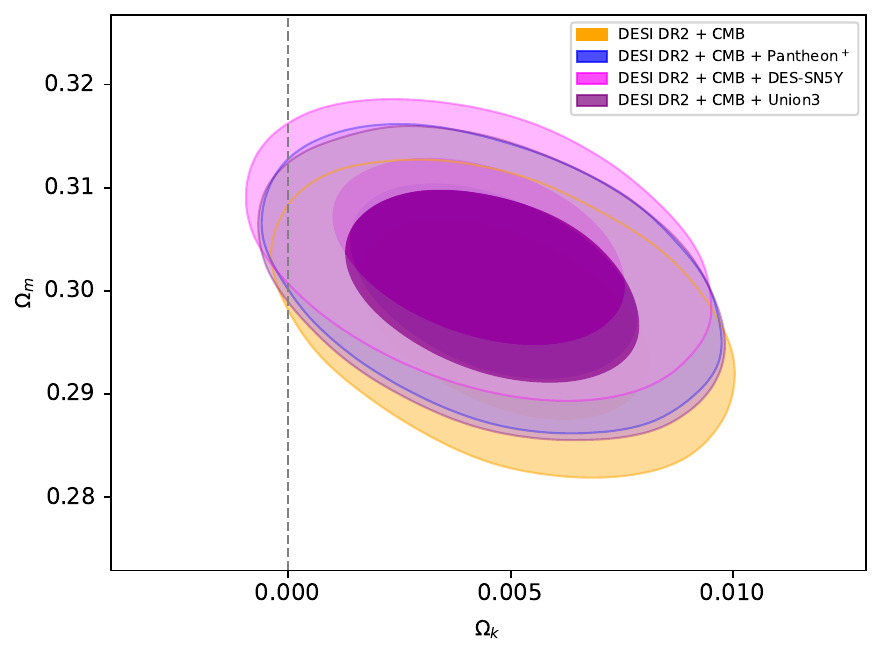}
    \caption{o$\Lambda$CDM}\label{fig_1a}
\end{subfigure}
\hfil
\begin{subfigure}{.3\textwidth}
\includegraphics[width=\linewidth]{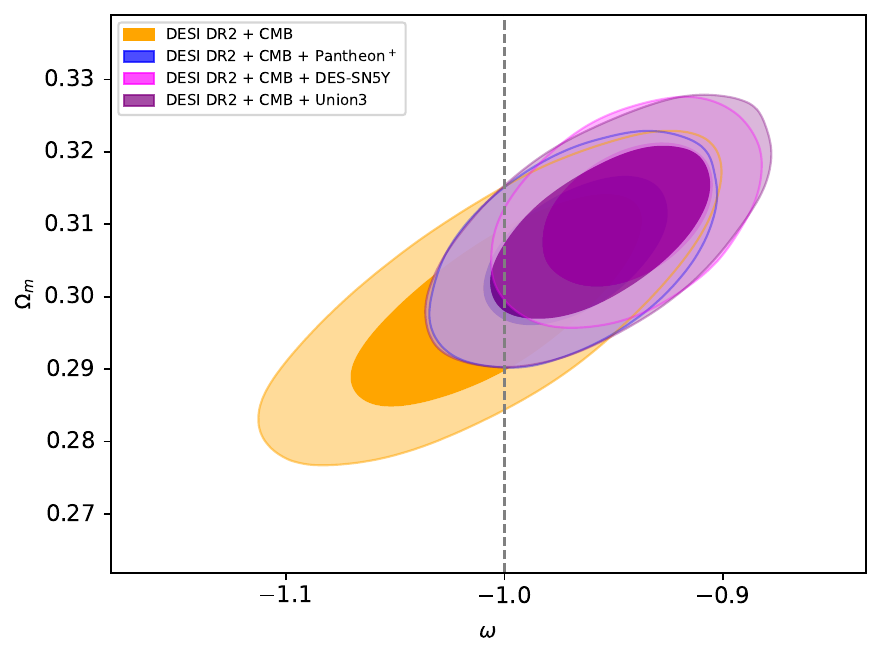}
    \caption{$\omega$CDM}\label{fig_1b}
\end{subfigure}
\hfil
\begin{subfigure}{.3\textwidth}
\includegraphics[width=\linewidth]{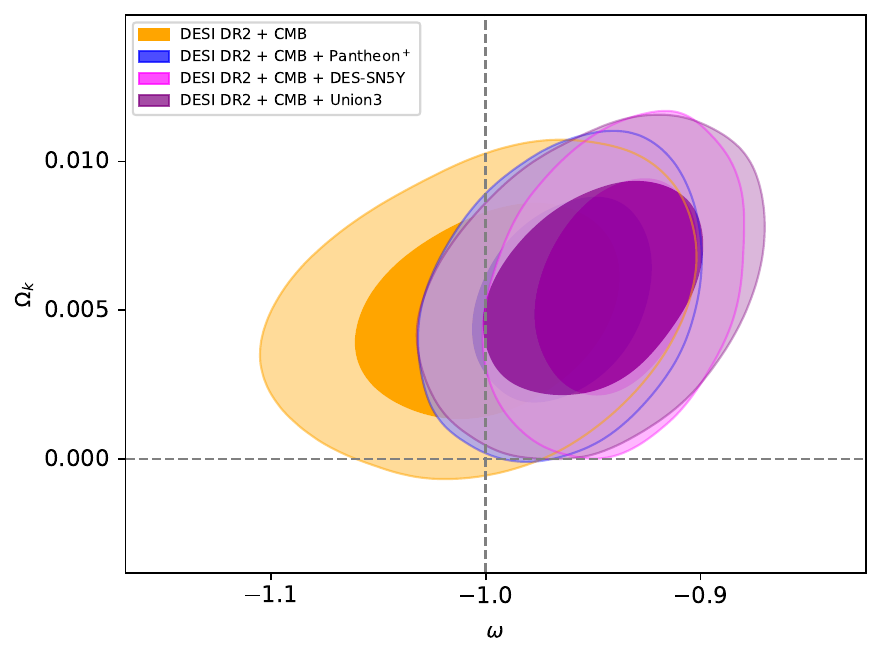}
    \caption{o$\omega$CDM}\label{fig_1c}
\end{subfigure}
\begin{subfigure}{.3\textwidth}
\includegraphics[width=\linewidth]{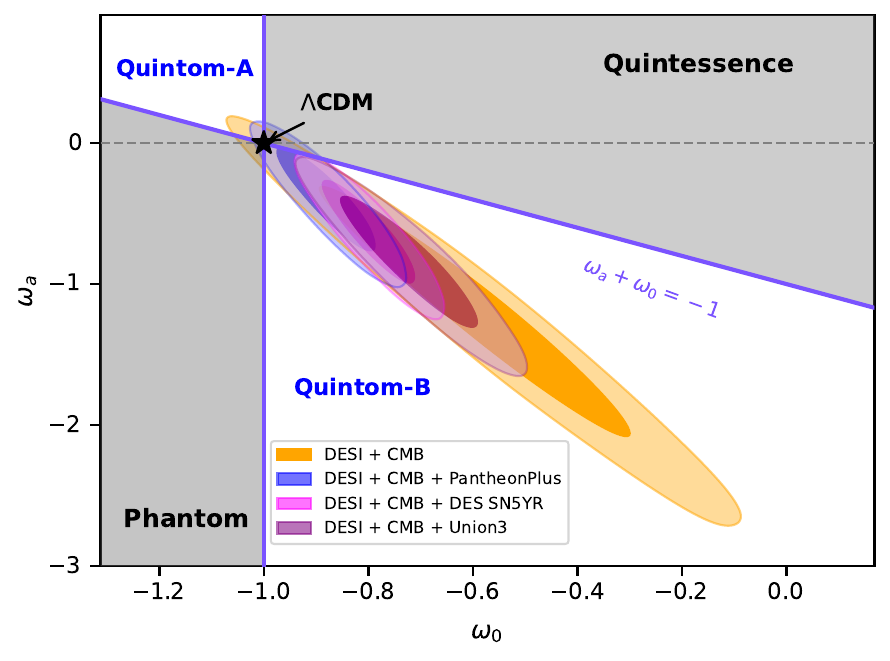}
    \caption{$\omega_0\omega_a$CDM}\label{fig_1d}
\end{subfigure}
\hfil
\begin{subfigure}{.3\textwidth}
\includegraphics[width=\linewidth]{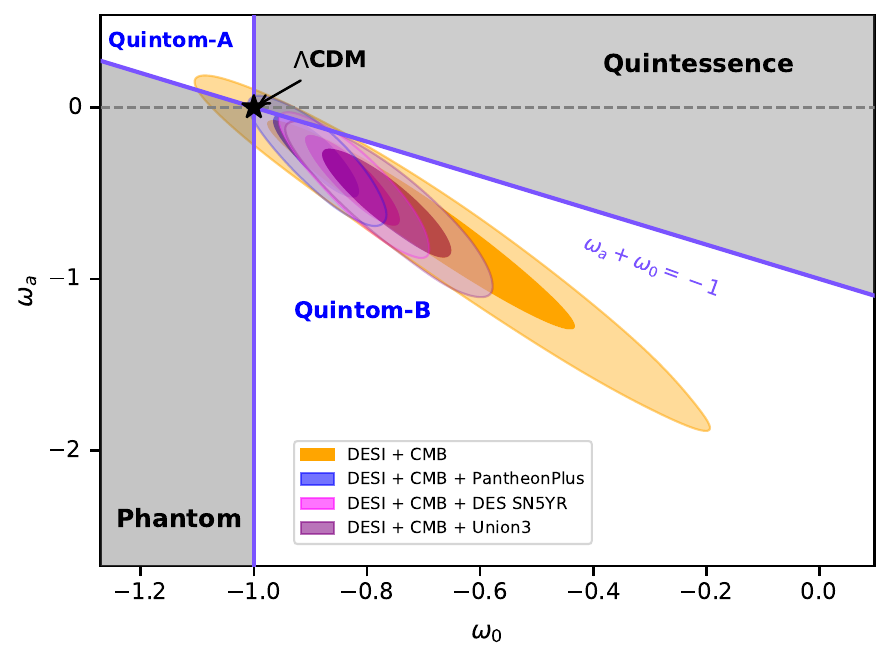}
     \caption{Logarithmic}\label{fig_1e}
\end{subfigure}
\hfil
\begin{subfigure}{.3\textwidth}
\includegraphics[width=\linewidth]{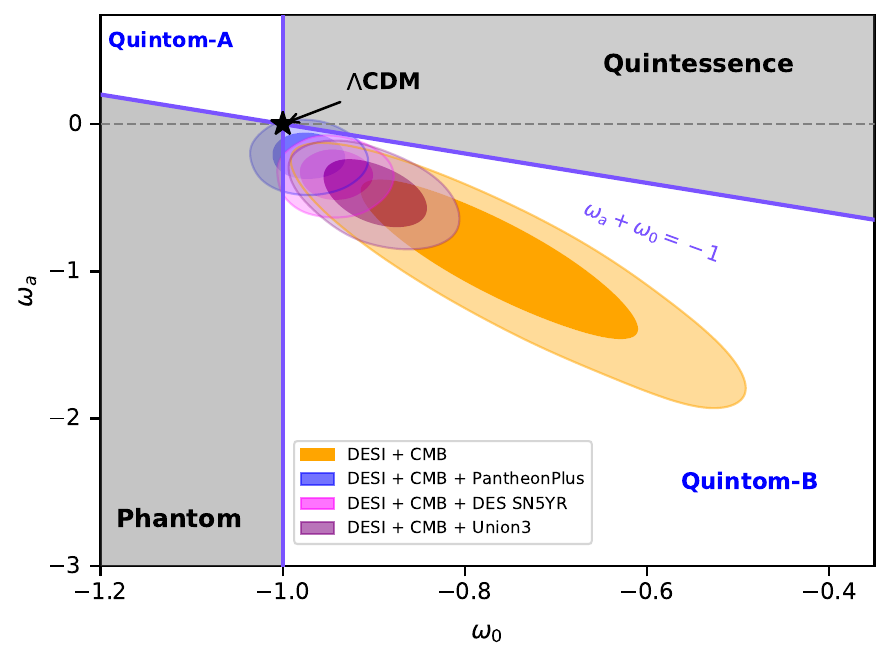}
    \caption{Exponential}\label{fig_1f}
\end{subfigure}
\hfil
\begin{subfigure}{.3\textwidth}
\includegraphics[width=\linewidth]{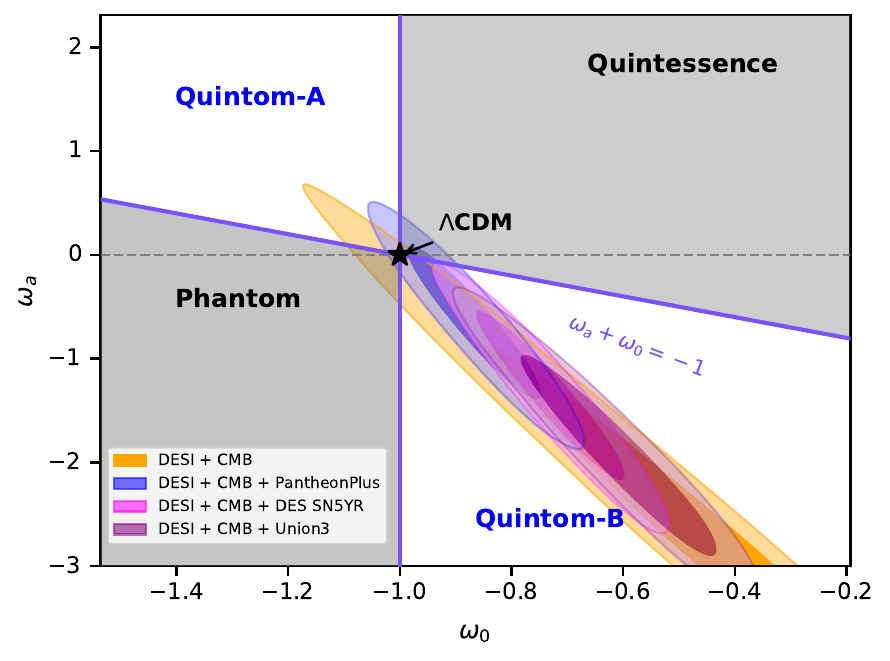}
    \caption{JBP}\label{fig_1g}
\end{subfigure}
\hfil
\begin{subfigure}{.3\textwidth}
\includegraphics[width=\linewidth]{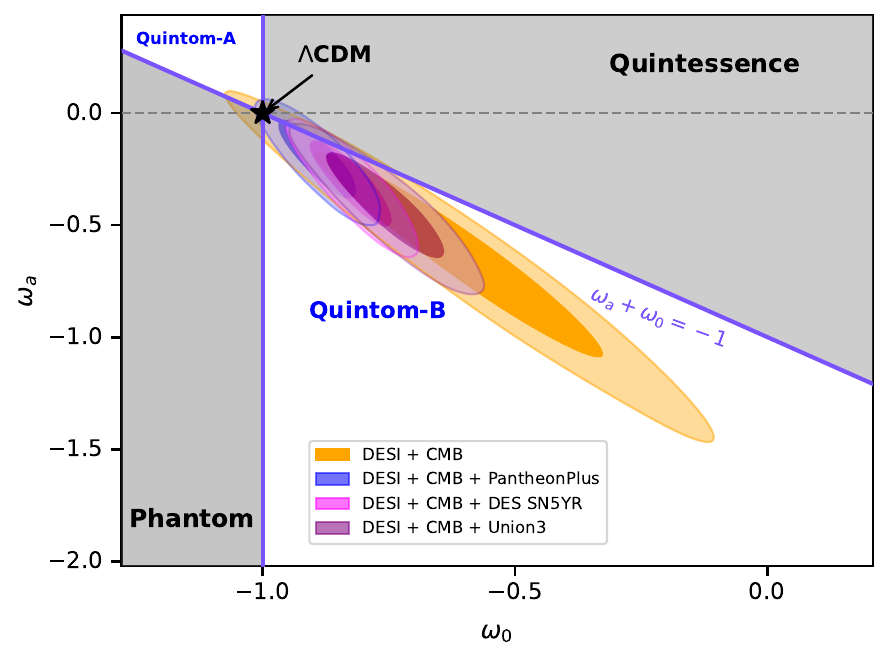}
     \caption{BA}\label{fig_1h}
\end{subfigure}
\hfil
\begin{subfigure}{.3\textwidth}
\includegraphics[width=\linewidth]{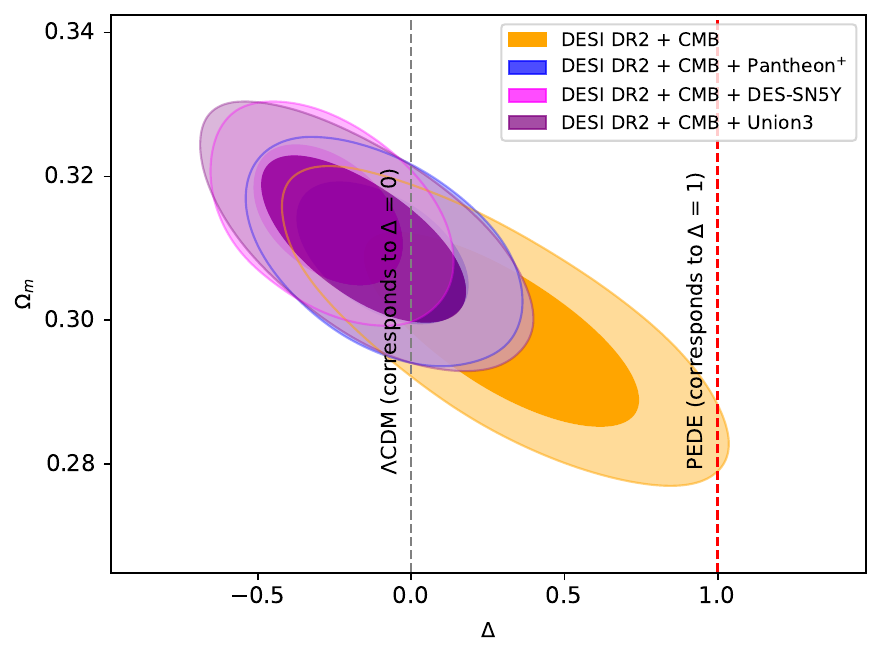}
     \caption{GEDE}\label{fig_1i}
\end{subfigure}
\caption{The figure shows the posterior distributions of different planes of the o$\Lambda$CDM, $\omega$CDM, o$\omega$CDM, $\omega_a \omega_0$CDM, Logarithmic, Exponential, JBP, BA, and GEDE models using DESI DR2 measurements in combination with the CMB and SNe Ia measurements, at the 68\% (1$\sigma$) and 95\% (2$\sigma$) confidence intervals.}\label{fig_1}
\end{figure*}

\begin{figure*}
\begin{subfigure}{.3\textwidth}
\includegraphics[width=\linewidth]{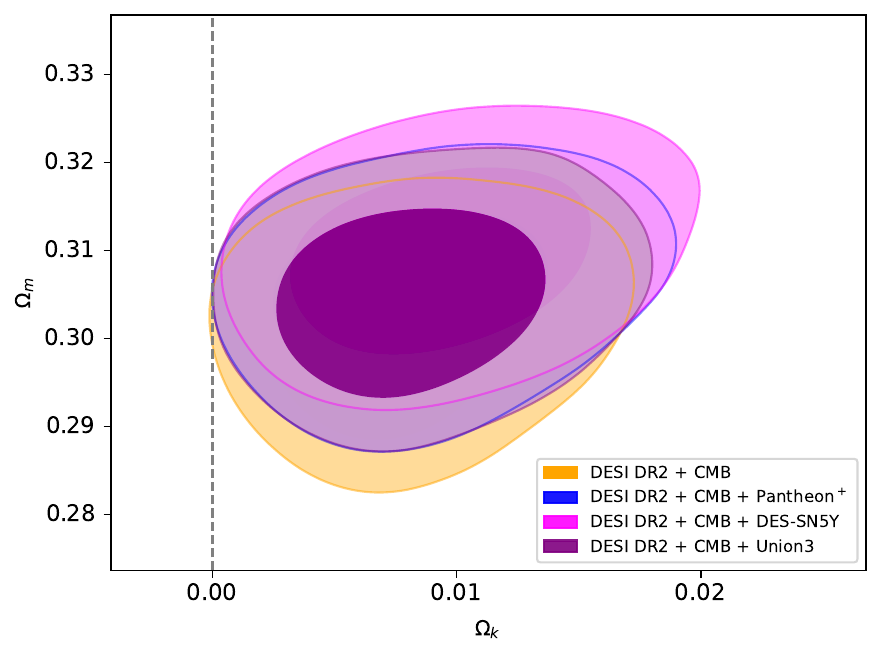}
    \caption{o$\Lambda$CDM + $\sum m_\nu $}\label{fig_2a}
\end{subfigure}
\hfil
\begin{subfigure}{.3\textwidth}
\includegraphics[width=\linewidth]{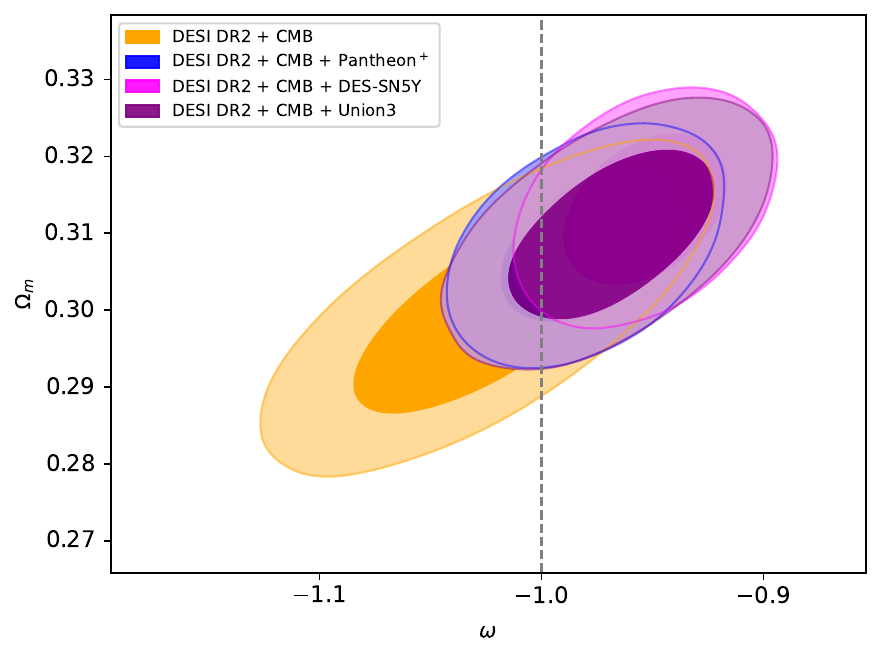}
    \caption{$\omega$CDM + $\sum m_\nu $}\label{fig_2b}
\end{subfigure}
\hfil
\begin{subfigure}{.3\textwidth}
\includegraphics[width=\linewidth]{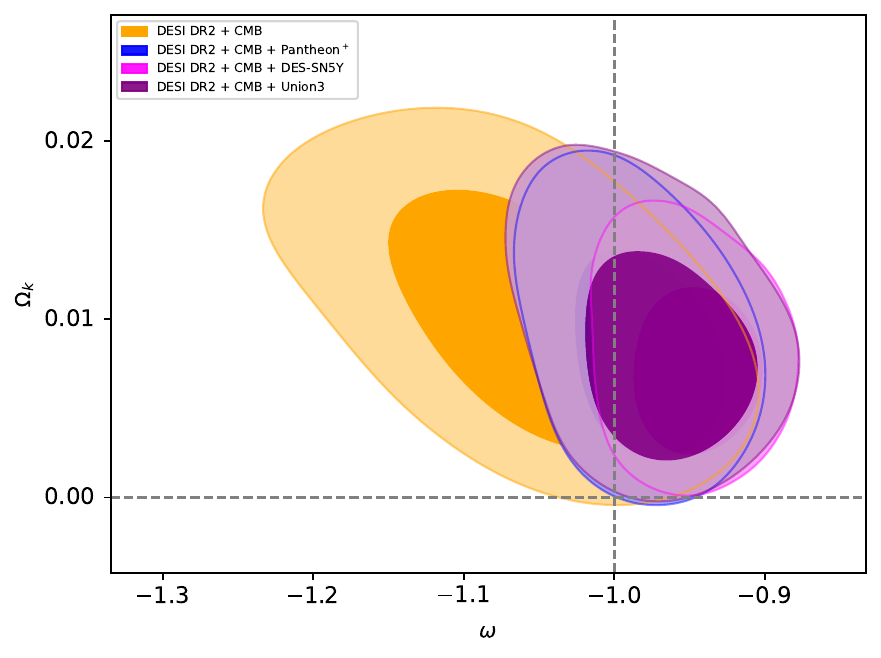}
    \caption{o$\omega$CDM + $\sum m_\nu $}\label{fig_2c}
\end{subfigure}
\begin{subfigure}{.3\textwidth}
\includegraphics[width=\linewidth]{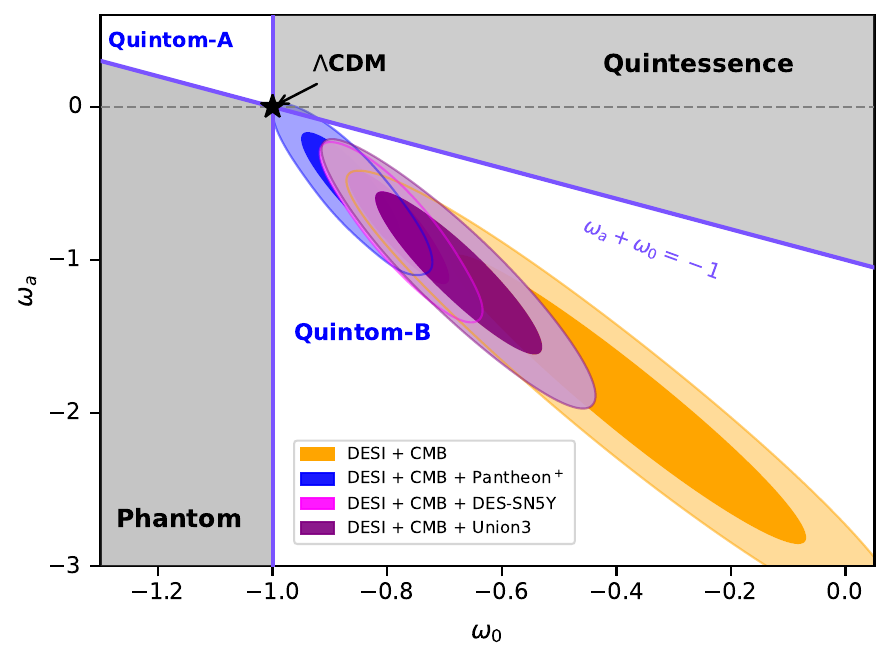}
    \caption{$\omega_0\omega_a$CDM + $\sum m_\nu $}\label{fig_2d}
\end{subfigure}
\hfil
\begin{subfigure}{.3\textwidth}
\includegraphics[width=\linewidth]{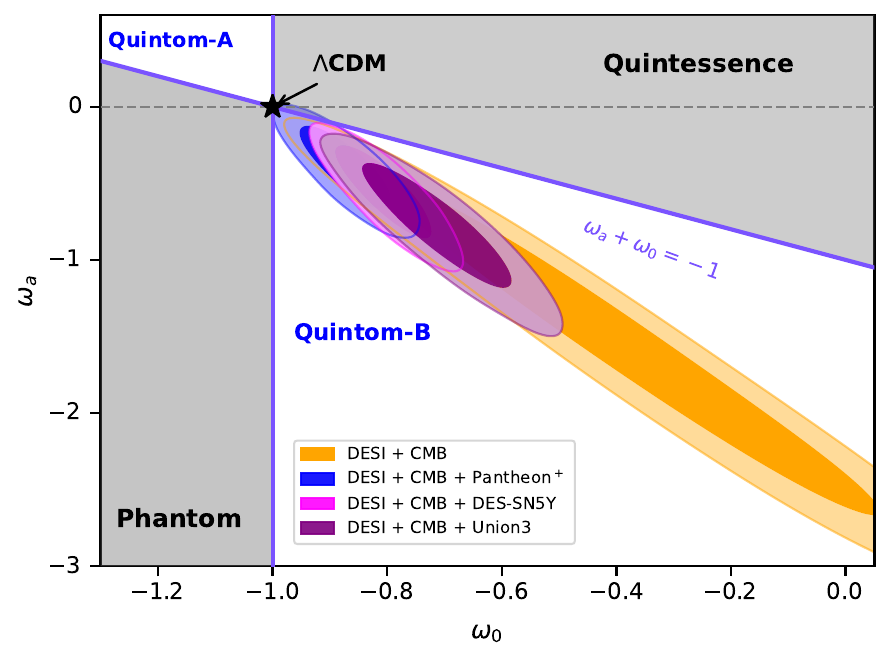}
     \caption{Logarithmic + $\sum m_\nu $}\label{fig_2e}
\end{subfigure}
\hfil
\begin{subfigure}{.3\textwidth}
\includegraphics[width=\linewidth]{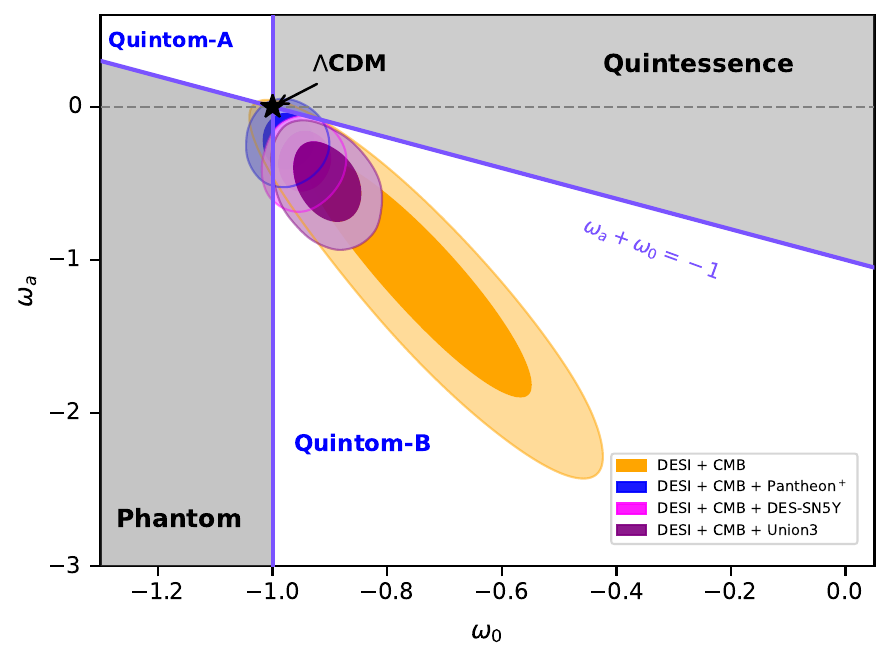}
    \caption{Exponential + $\sum m_\nu $}\label{fig_2f}
\end{subfigure}
\hfil
\begin{subfigure}{.3\textwidth}
\includegraphics[width=\linewidth]{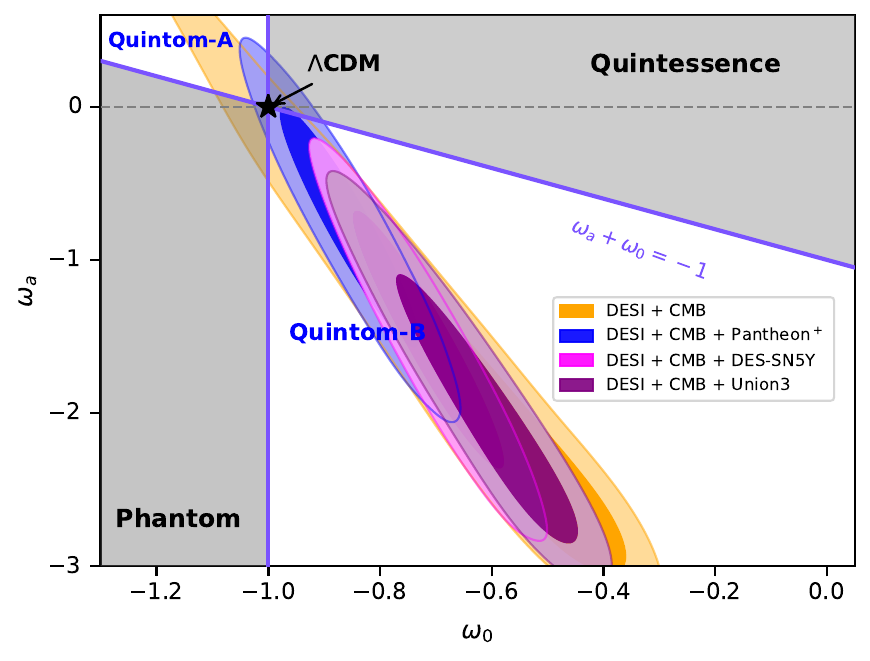}
    \caption{JBP + $\sum m_\nu $}\label{fig_2g}
\end{subfigure}
\hfil
\begin{subfigure}{.3\textwidth}
\includegraphics[width=\linewidth]{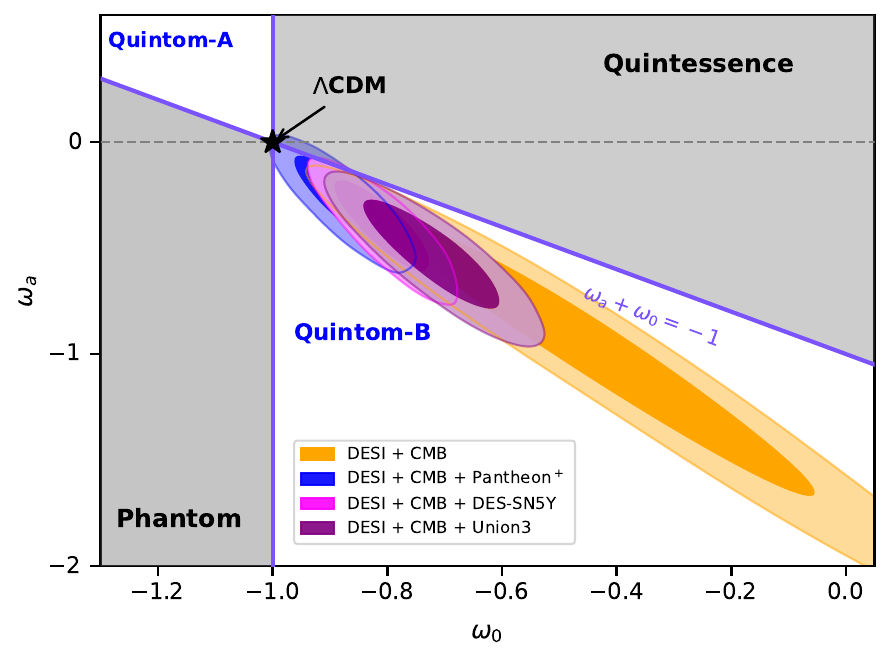}
     \caption{BA + $\sum m_\nu $}\label{fig_2h}
\end{subfigure}
\hfil
\begin{subfigure}{.3\textwidth}
\includegraphics[width=\linewidth]{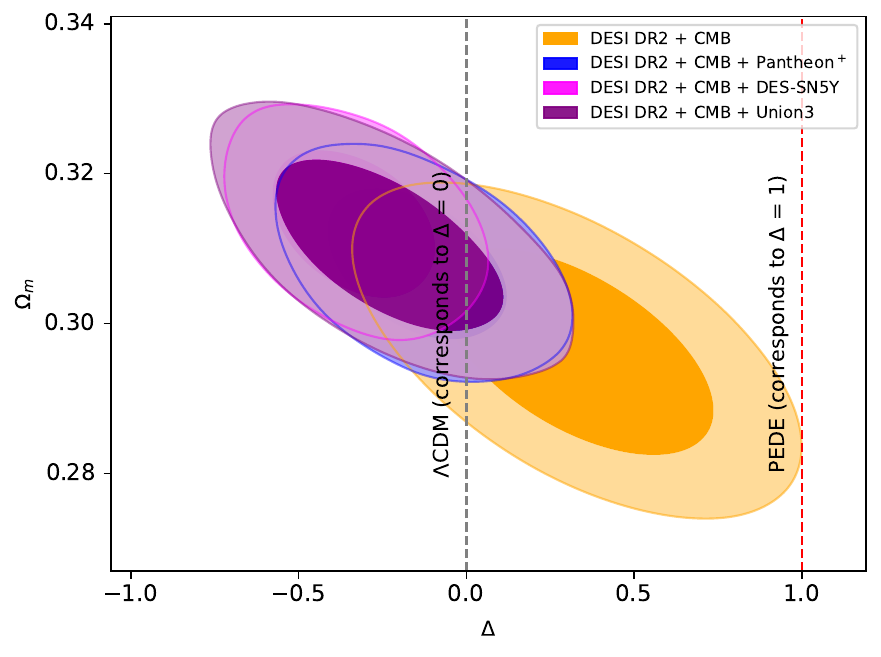}
     \caption{GEDE + $\sum m_\nu $}\label{fig_2i}
\end{subfigure}
\caption{The figure shows the posterior distributions of different planes of the o$\Lambda$CDM + $\sum m_\nu $, $\omega$CDM + $\sum m_\nu $, o$\omega$CDM + $\sum m_\nu $, $\omega_a \omega_0$CDM + $\sum m_\nu $, Logarithmic + $\sum m_\nu $, Exponential + $\sum m_\nu $, JBP + $\sum m_\nu $, BA + $\sum m_\nu $, and GEDE + $\sum m_\nu $ models using DESI DR2 measurements in combination with the CMB and SNe Ia measurements, at the 68\% (1$\sigma$) and 95\% (2$\sigma$) confidence intervals.}\label{fig_2}
\end{figure*}

\begin{figure}
\centering
\includegraphics[scale=0.54]{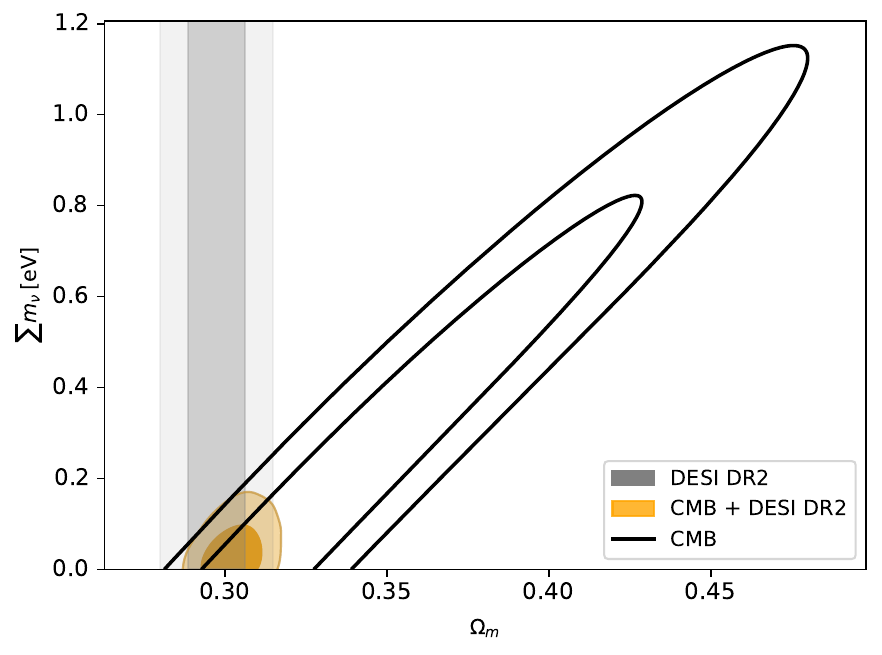}
\caption{The figure shows the contour plot in the $\Omega_m$–$\sum m_\nu$ plane for the $\Lambda$CDM model, using the CMB alone and combined with the DESI DR2 dataset, showing the 68\% and 95\% confidence level contours.}\label{fig_sum_LCDM}
\end{figure}

\begin{figure*}
\begin{subfigure}{.27\textwidth}
\includegraphics[width=\linewidth]{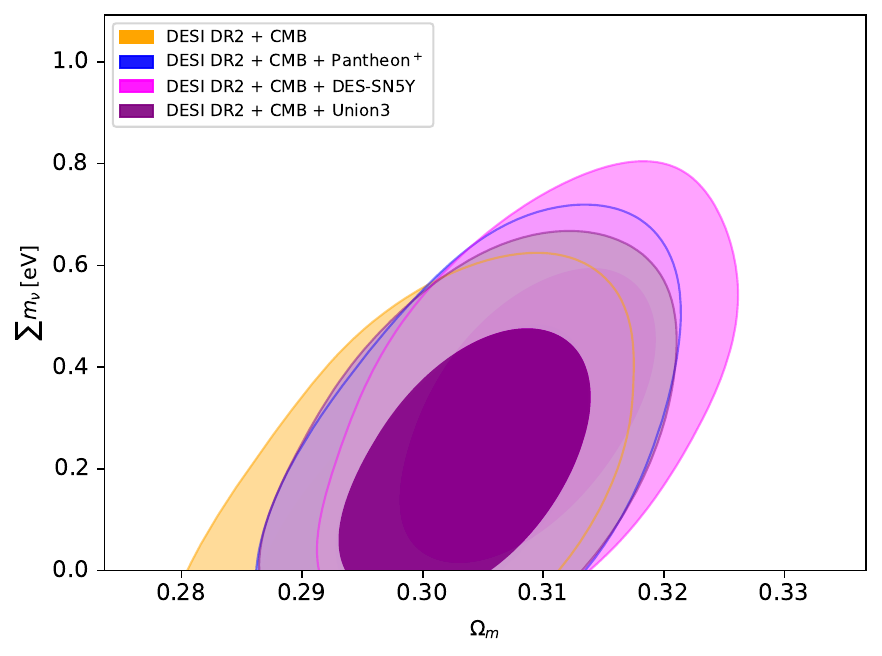}
    \caption{o$\Lambda$CDM + $\sum m_\nu $}\label{fig_3a}
\end{subfigure}
\hfil
\begin{subfigure}{.27\textwidth}
\includegraphics[width=\linewidth]{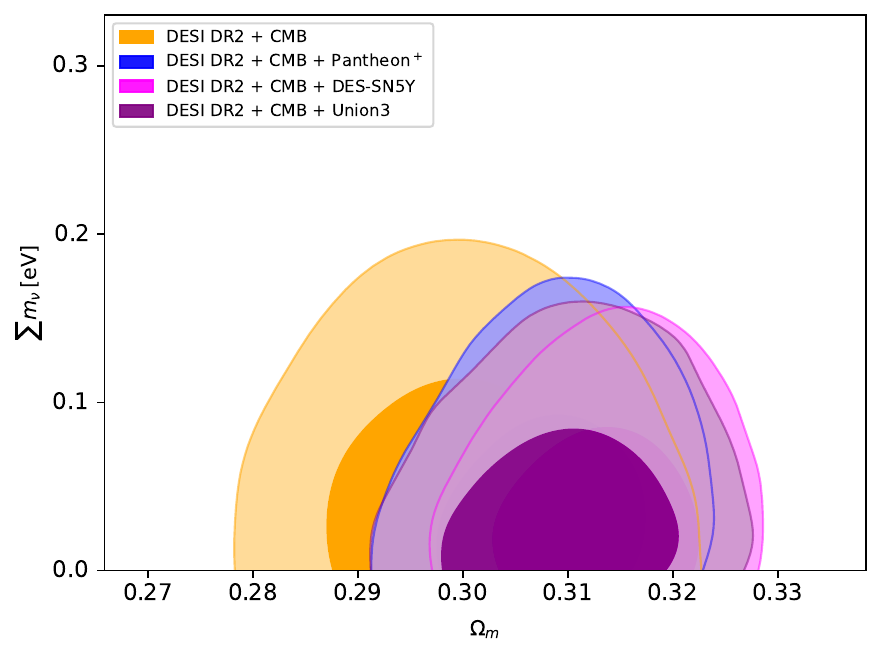}
    \caption{$\omega$CDM + $\sum m_\nu $}\label{fig_3b}
\end{subfigure}
\hfil
\begin{subfigure}{.27\textwidth}
\includegraphics[width=\linewidth]{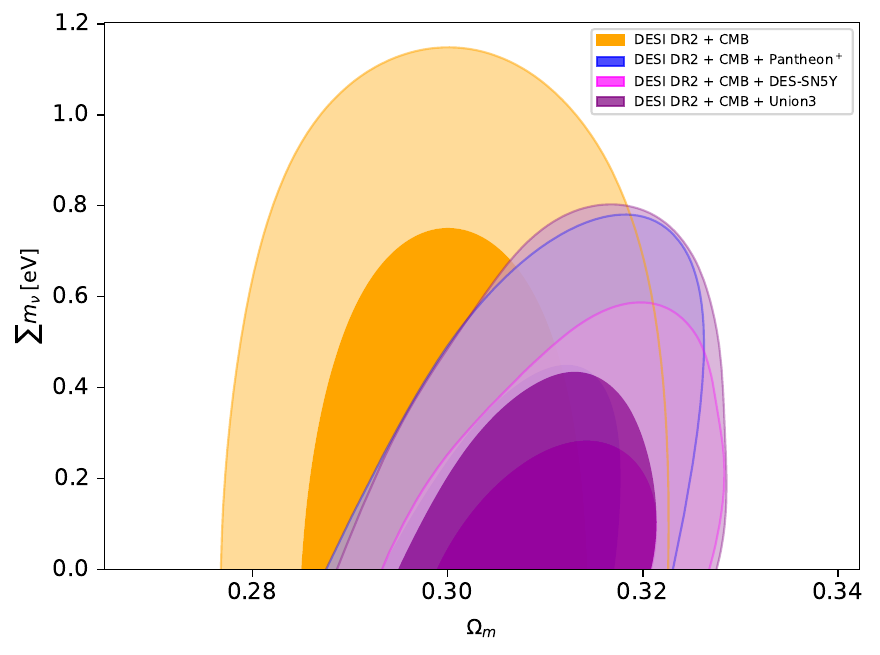}
    \caption{o$\omega$CDM + $\sum m_\nu $}\label{fig_3c}
\end{subfigure}
\begin{subfigure}{.30\textwidth}
\includegraphics[width=\linewidth]{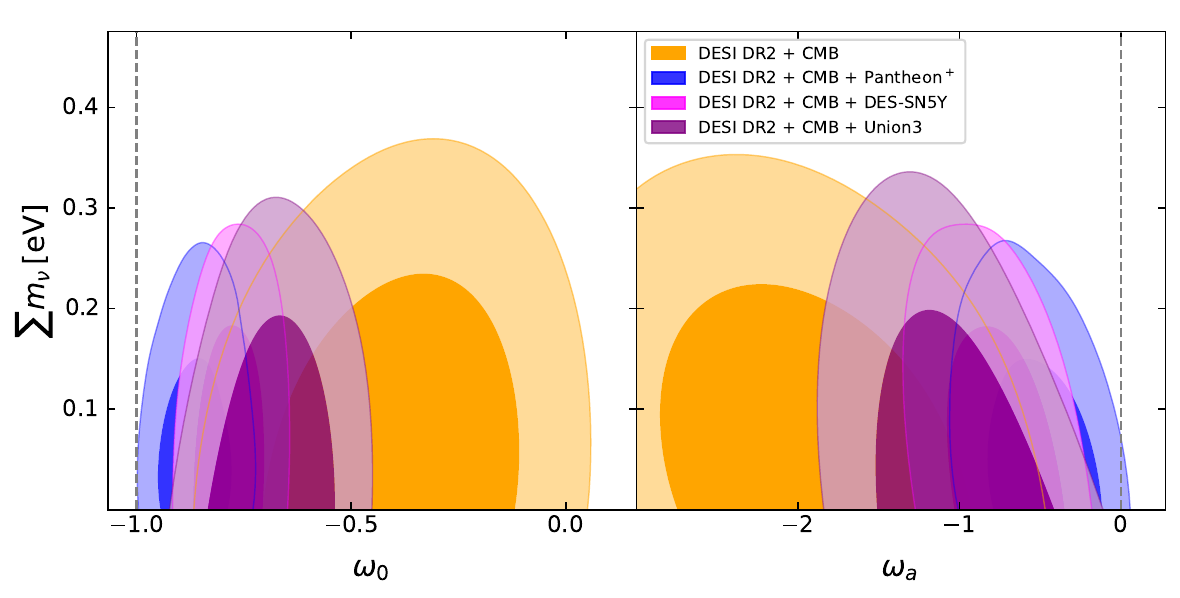}
    \caption{$\omega_0\omega_a$CDM + $\sum m_\nu $}\label{fig_3d}
\end{subfigure}
\hfil
\begin{subfigure}{.30\textwidth}
\includegraphics[width=\linewidth]{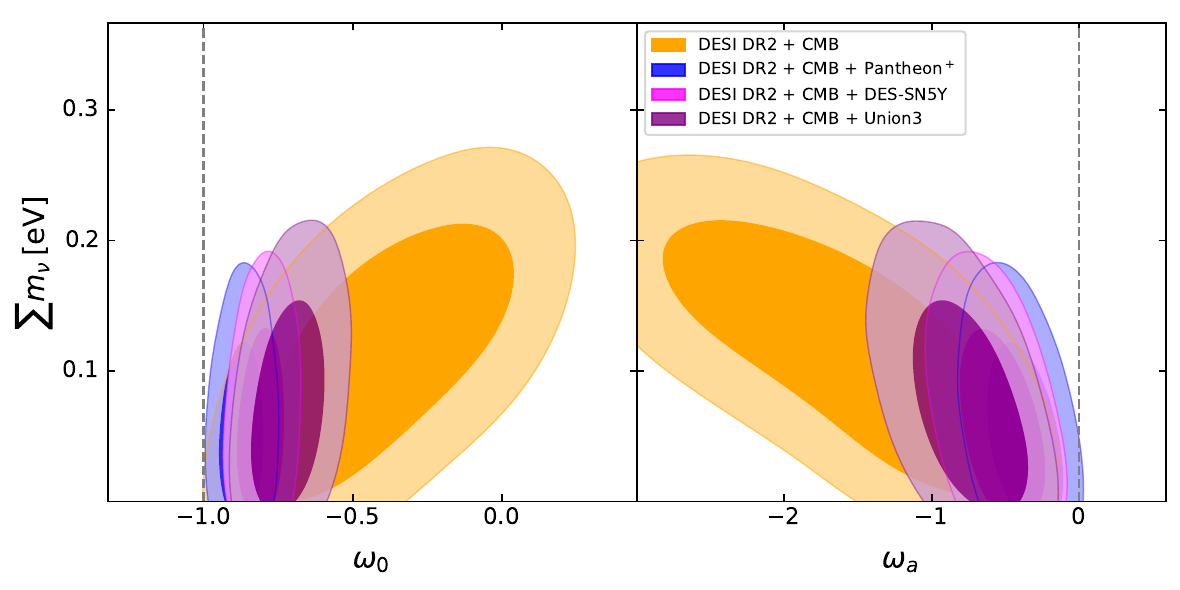}
     \caption{Logarithmic + $\sum m_\nu $}\label{fig_3e}
\end{subfigure}
\hfil
\begin{subfigure}{.30\textwidth}
\includegraphics[width=\linewidth]{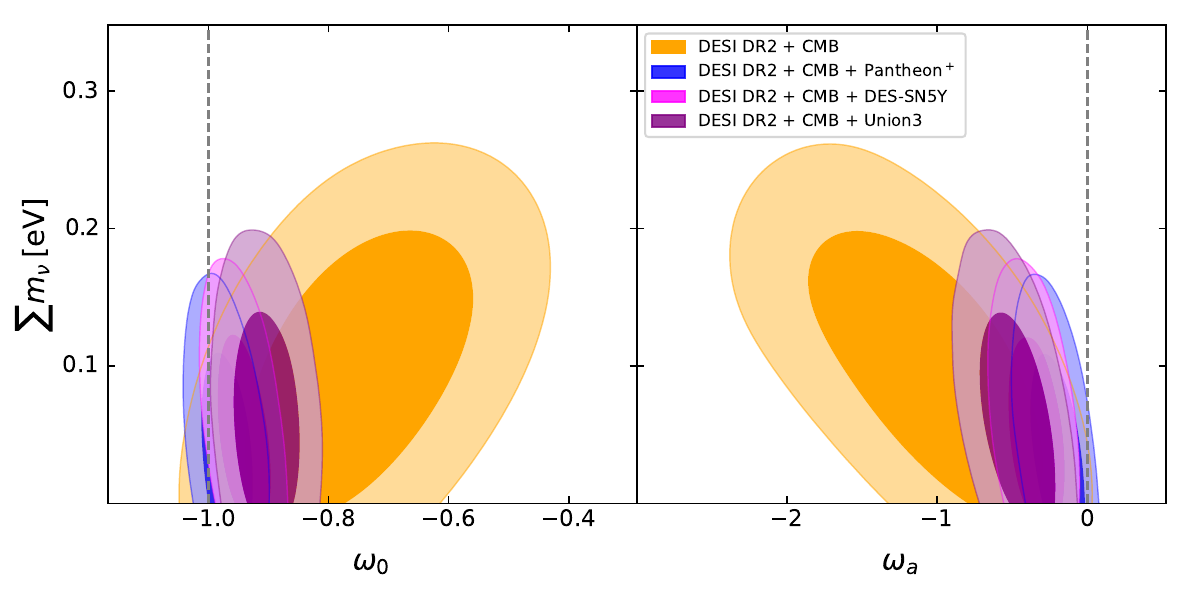}
    \caption{Exponential + $\sum m_\nu $}\label{fig_3f}
\end{subfigure}
\hfil
\begin{subfigure}{.32\textwidth}
\includegraphics[width=\linewidth]{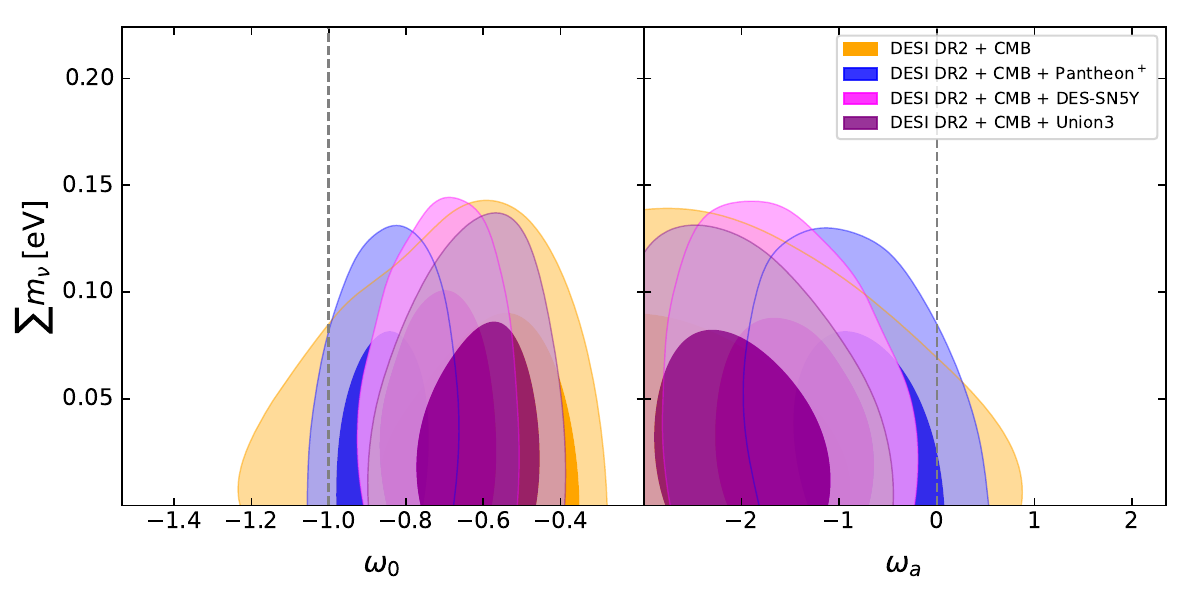}
    \caption{JBP + $\sum m_\nu $}\label{fig_3g}
\end{subfigure}
\hfil
\begin{subfigure}{.32\textwidth}
\includegraphics[width=\linewidth]{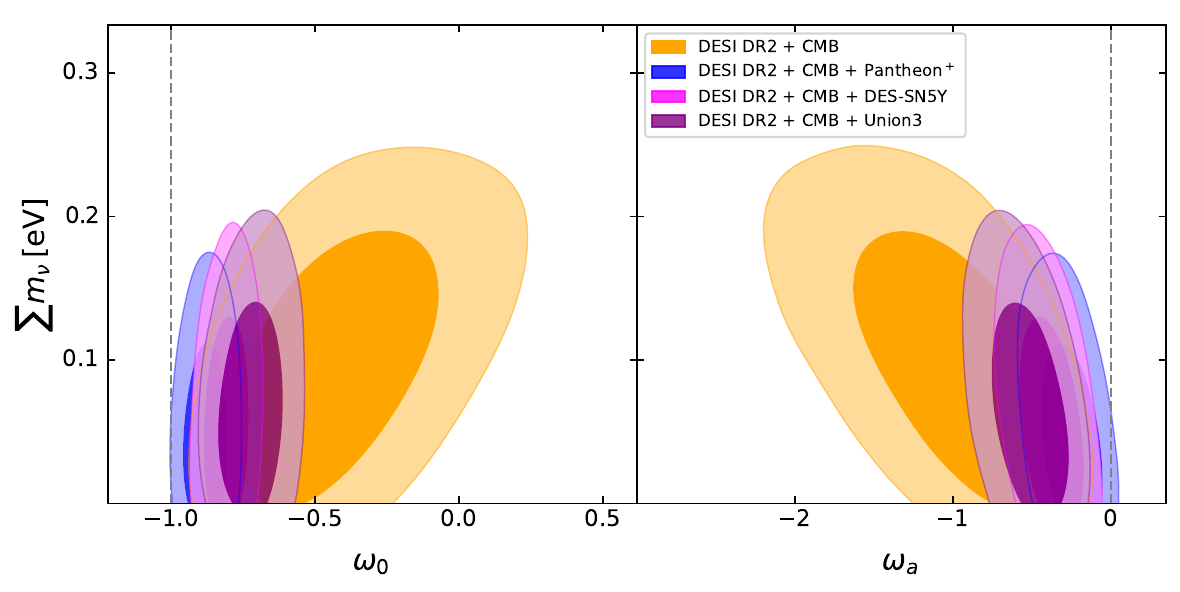}
     \caption{BA + $\sum m_\nu $}\label{fig_3h}
\end{subfigure}
\hfil
\begin{subfigure}{.27\textwidth}
\includegraphics[width=\linewidth]{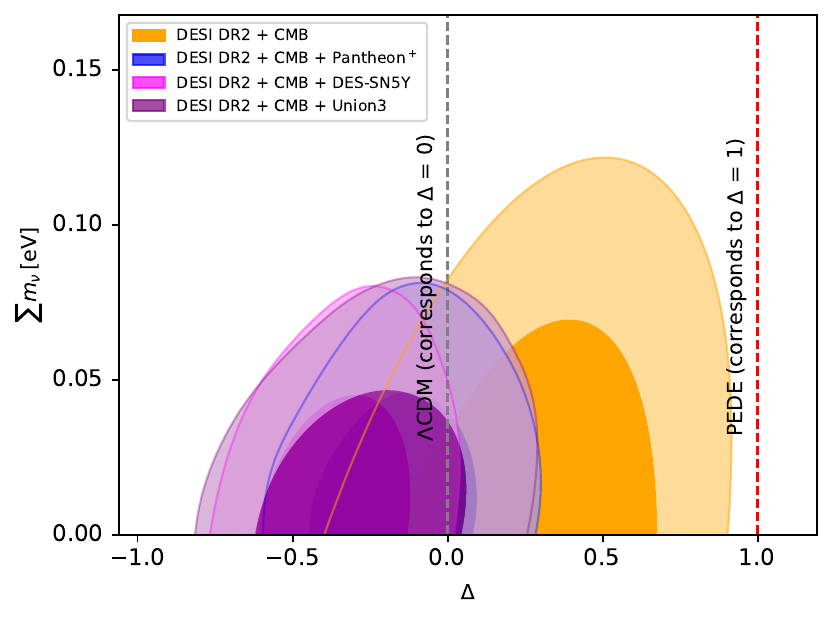}
     \caption{GEDE + $\sum m_\nu $}\label{fig_3i}
\end{subfigure}
\caption{The figure shows the posterior distributions different planes of the o$\Lambda$CDM + $\sum m_\nu $, $\omega$CDM + $\sum m_\nu $, o$\omega$CDM + $\sum m_\nu $, $\omega_a \omega_0$CDM + $\sum m_\nu $, Logarithmic + $\sum m_\nu $, Exponential + $\sum m_\nu $, JBP + $\sum m_\nu $, BA + $\sum m_\nu $, and GEDE + $\sum m_\nu$ models using DESI DR2 measurements in combination with the CMB and SNe Ia measurements, at the 68\% (1$\sigma$) and 95\% (2$\sigma$) confidence intervals.}\label{fig_3}
\end{figure*}

\begin{figure}
\centering
\includegraphics[scale=0.52]{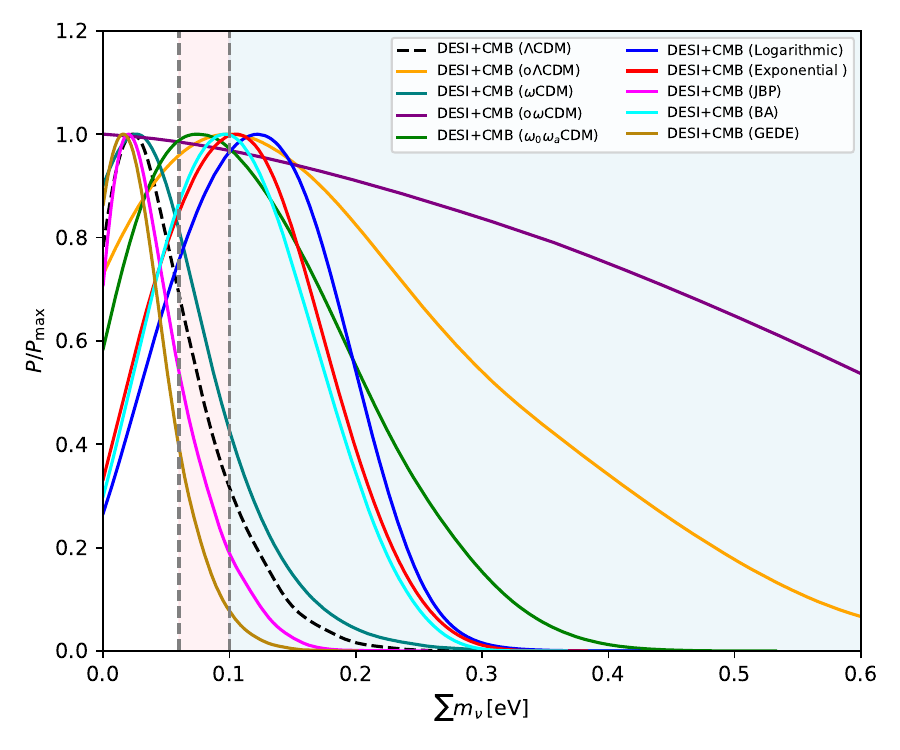}
\caption{The figure shows the 1D marginalized posterior constraints on $\sum m_\nu$ for each model considered in the paper, using the combination of DESI DR2 and CMB}\label{fig_posterior}
\end{figure}
\begin{figure*}
\begin{subfigure}{.3\textwidth}
\includegraphics[width=\linewidth]{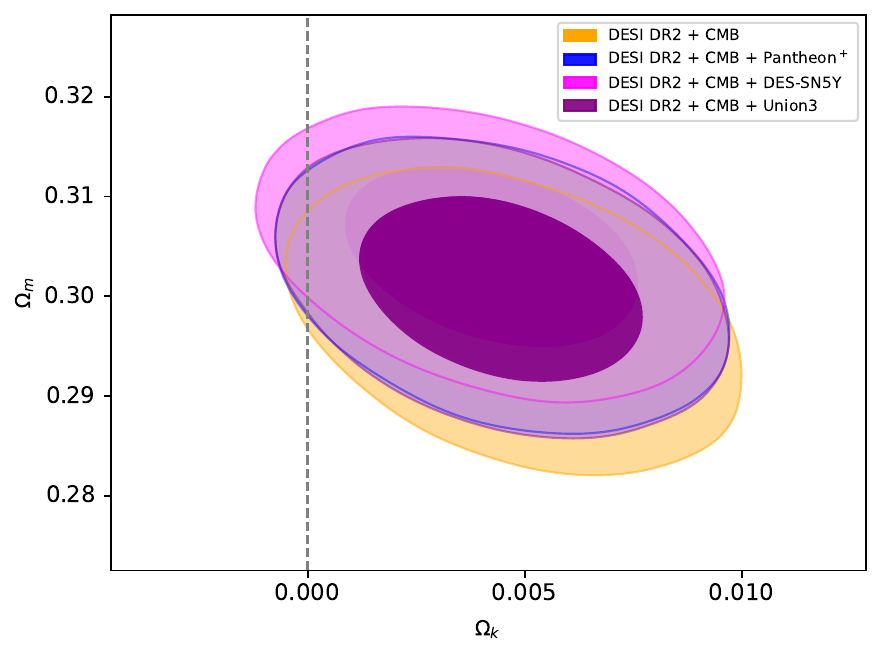}
    \caption{o$\Lambda$CDM + $N_{eff}$}\label{fig_4a}
\end{subfigure}
\hfil
\begin{subfigure}{.3\textwidth}
\includegraphics[width=\linewidth]{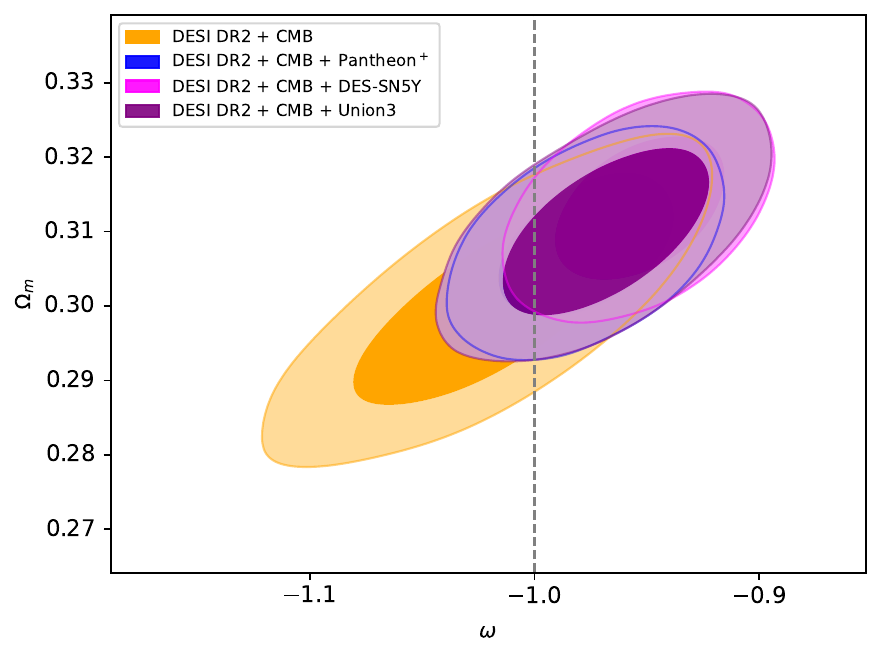}
    \caption{$\omega$CDM + $N_{eff} $}\label{fig_4b}
\end{subfigure}
\hfil
\begin{subfigure}{.3\textwidth}
\includegraphics[width=\linewidth]{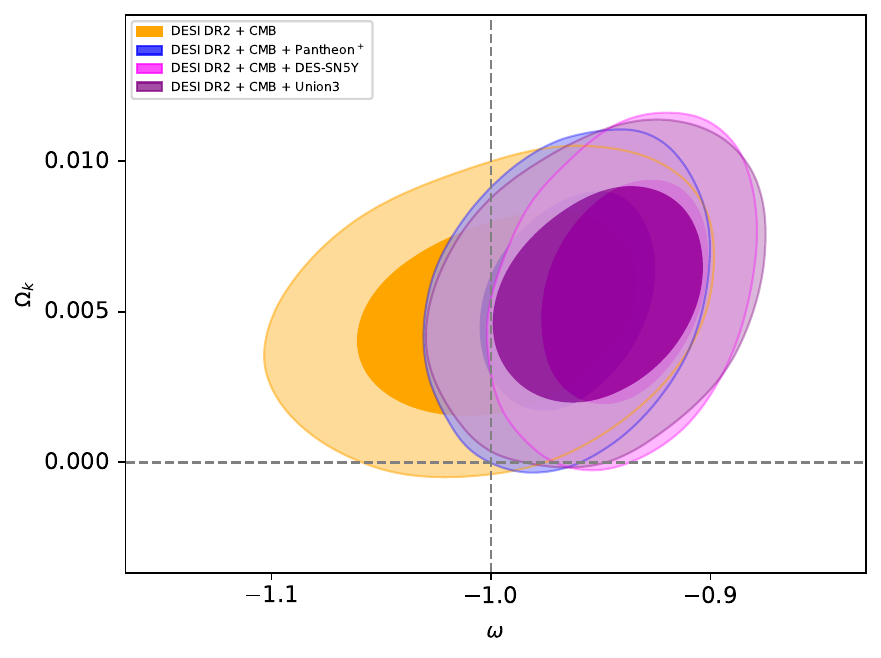}
    \caption{o$\omega$CDM + $N_{eff} $}\label{fig_4c}
\end{subfigure}
\begin{subfigure}{.3\textwidth}
\includegraphics[width=\linewidth]{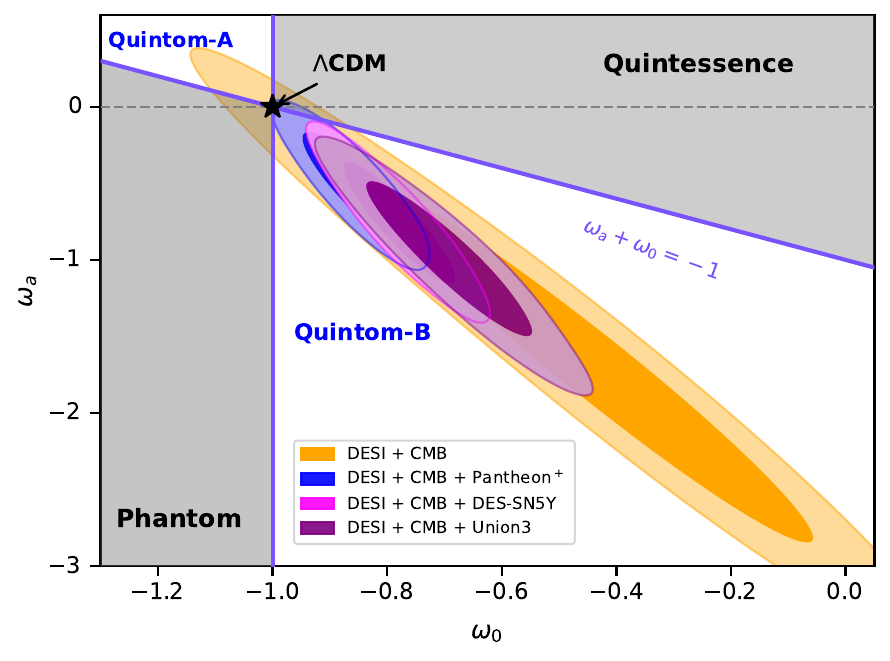}
    \caption{$\omega_0\omega_a$CDM + $N_{eff} $}\label{fig_4d}
\end{subfigure}
\hfil
\begin{subfigure}{.3\textwidth}
\includegraphics[width=\linewidth]{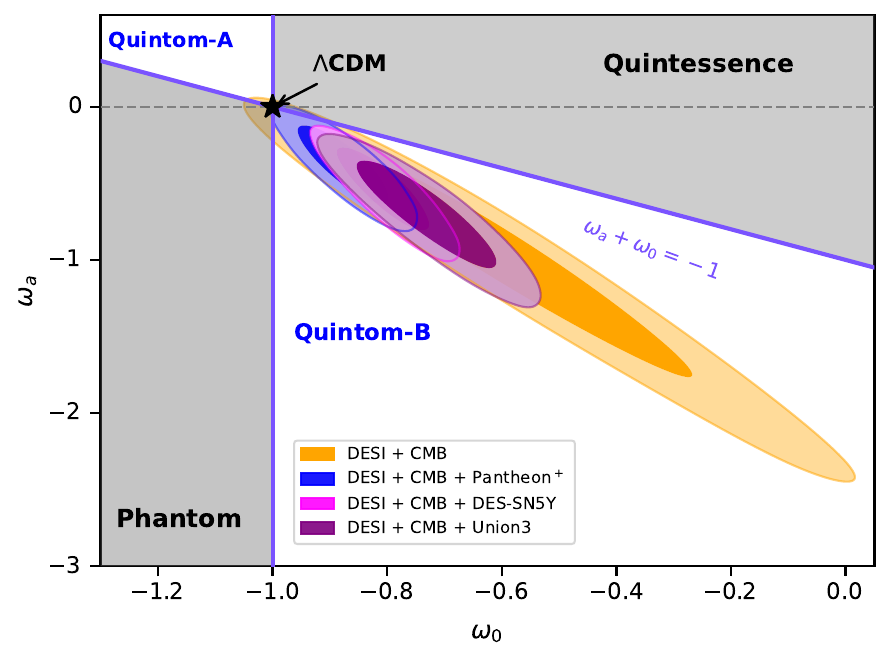}
     \caption{Logarithmic + $N_{eff} $}\label{fig_4e}
\end{subfigure}
\hfil
\begin{subfigure}{.3\textwidth}
\includegraphics[width=\linewidth]{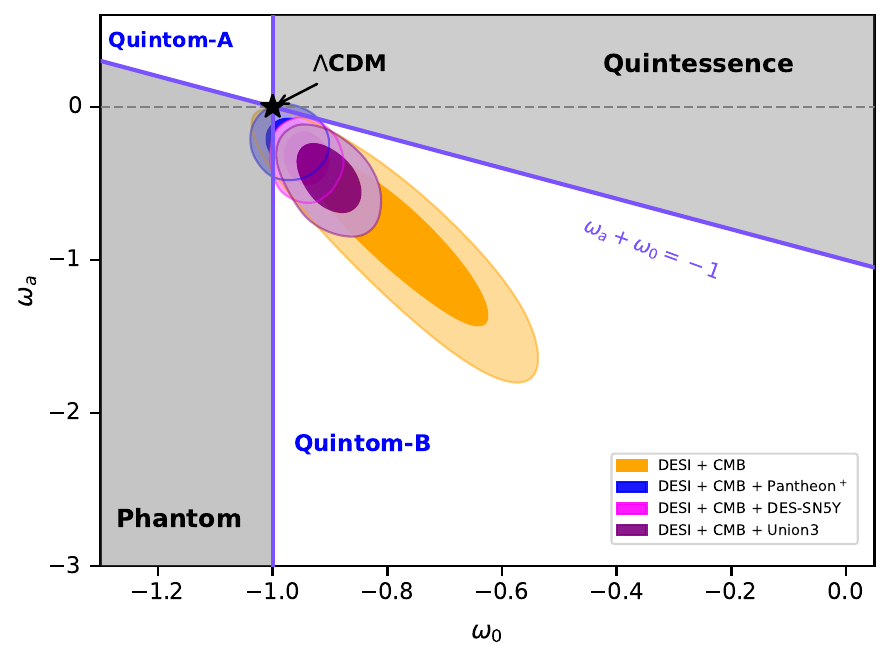}
    \caption{Exponential + $N_{eff} $}\label{fig_4f}
\end{subfigure}
\hfil
\begin{subfigure}{.3\textwidth}
\includegraphics[width=\linewidth]{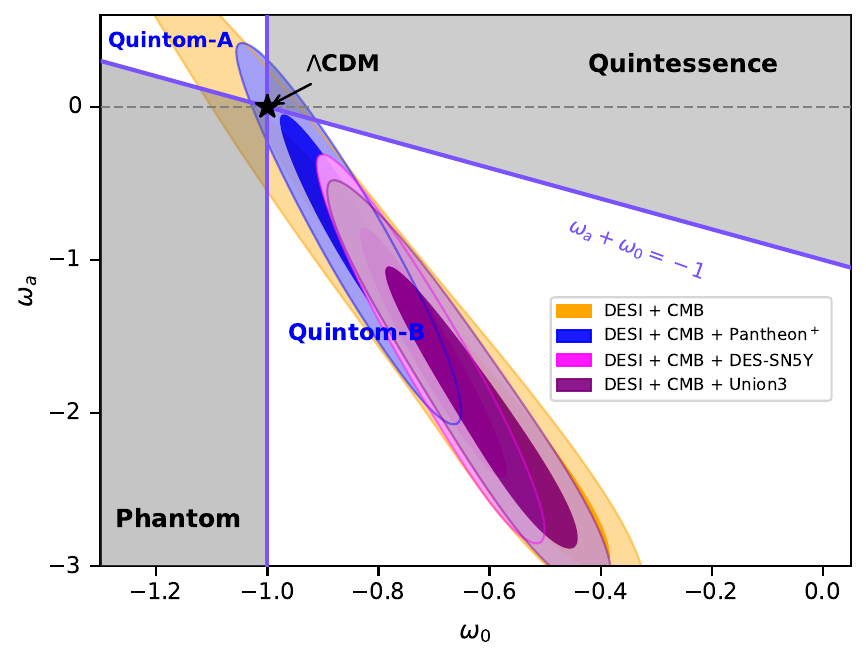}
    \caption{JBP + $N_{eff} $}\label{fig_4g}
\end{subfigure}
\hfil
\begin{subfigure}{.3\textwidth}
\includegraphics[width=\linewidth]{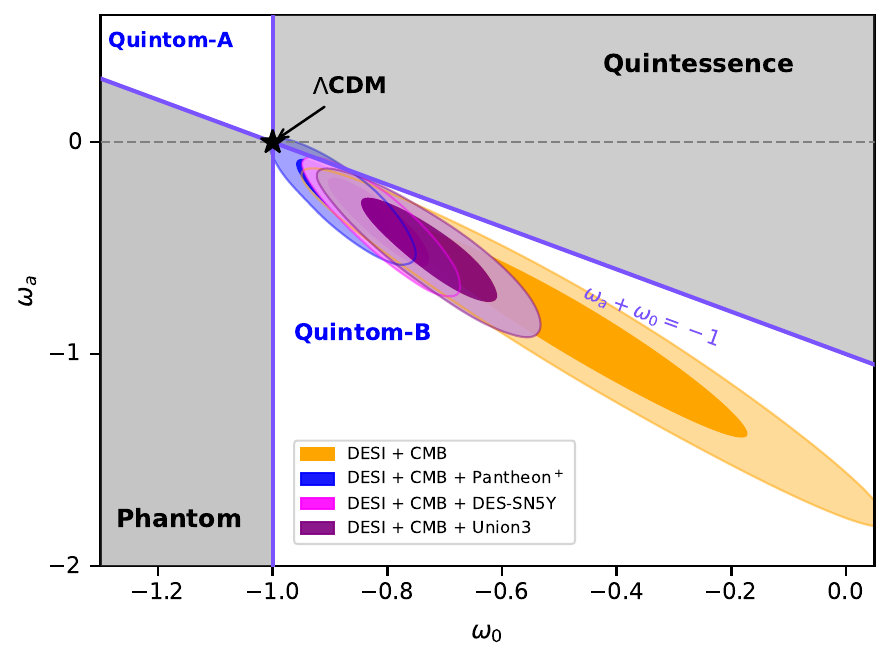}
     \caption{BA + $N_{eff} $}\label{fig_4h}
\end{subfigure}
\hfil
\begin{subfigure}{.3\textwidth}
\includegraphics[width=\linewidth]{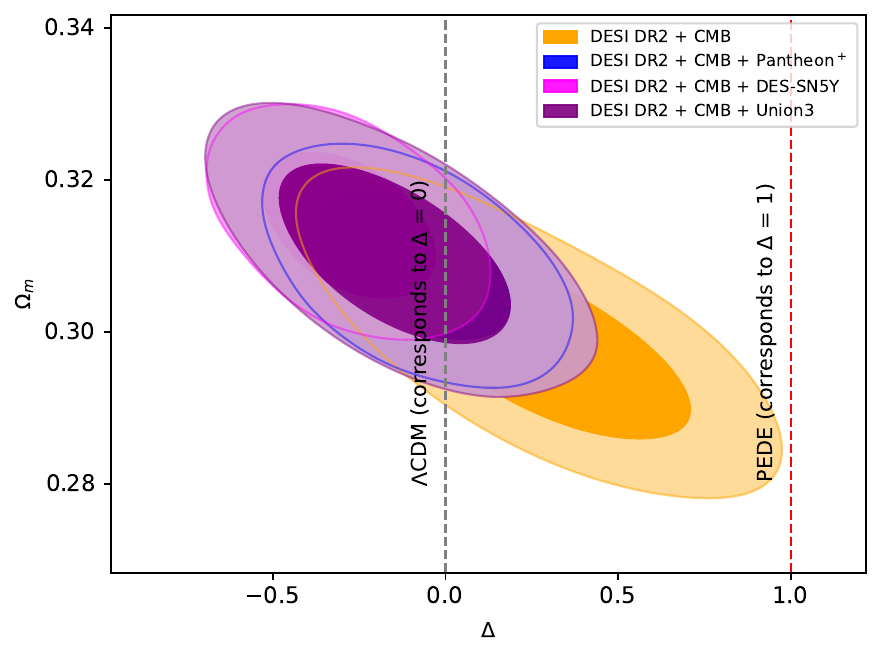}
     \caption{GEDE + $N_{eff} $}\label{fig_4i}
\end{subfigure}
\caption{The figure shows the posterior distributions of different planes of the o$\Lambda$CDM + $N_{eff} $, $\omega$CDM + $N_{eff} $, o$\omega$CDM + $N_{eff} $, $\omega_a \omega_0$CDM + $N_{eff} $, Logarithmic + $N_{eff} $, Exponential + $N_{eff} $, JBP + $N_{eff} $, BA + $N_{eff} $, and GEDE + $N_{eff} $ models using DESI DR2 measurements in combination with the CMB and SNe Ia measurements, at the 68\% (1$\sigma$) and 95\% (2$\sigma$) confidence intervals.}\label{fig_4}
\end{figure*}
\begin{figure*}
\centering
\includegraphics[scale=0.24]{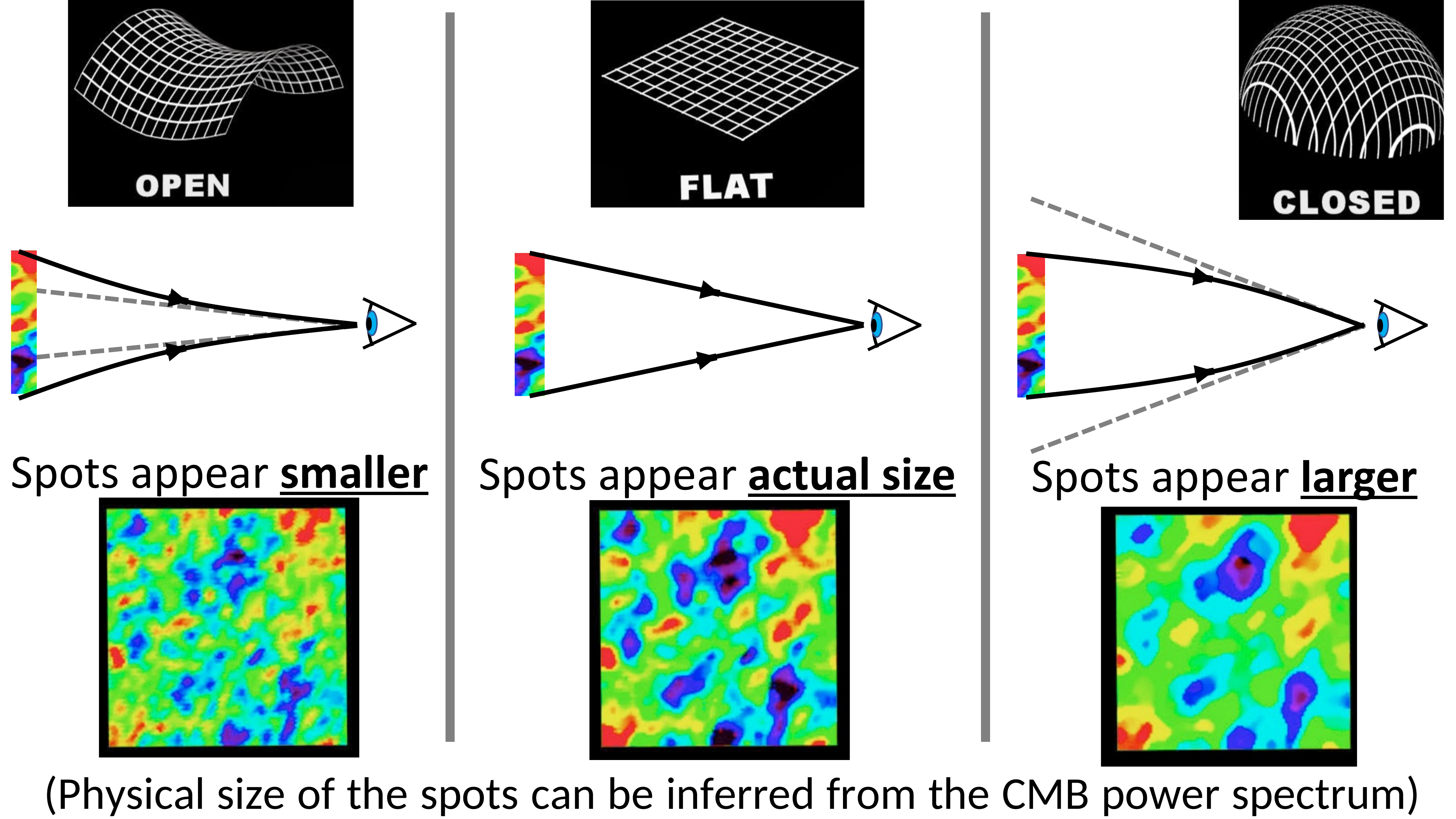}
\caption{The figure show the effect of spatial curvature on the apparent size of cosmic microwave background spots. Left: Open Universe  spots appear smaller due to negative curvature. Center: Flat Universe spots appear at their actual size. Right: Closed Universe spots appear larger due to positive curvature. The physical size of the spots can be deduced from the CMB power spectrum.}\label{fig_cmb}
\end{figure*}
\begin{figure*}
\centering
\includegraphics[scale=0.40]{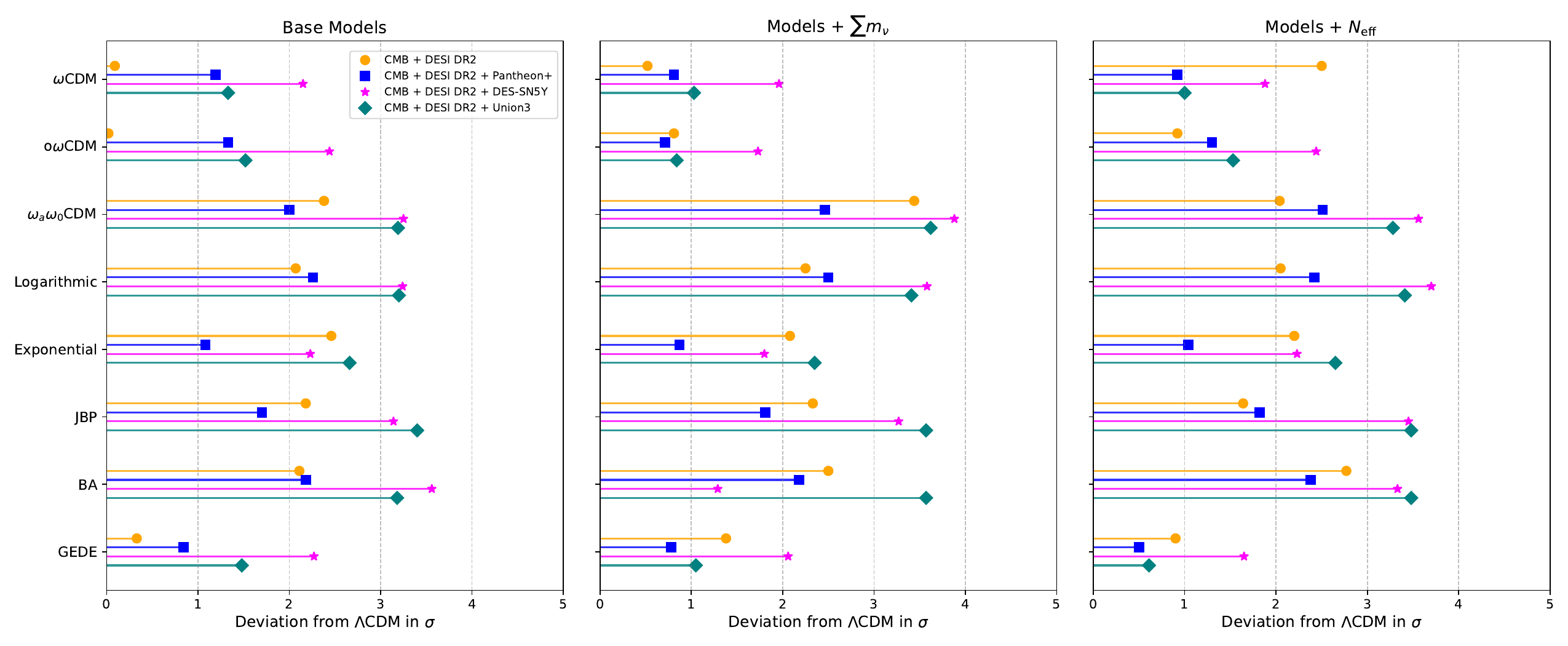}
\caption{The figure shows the deviation of each model from the $\Lambda$CDM model in the first column without varying $\sum m_\nu$ and $N_\mathrm{eff}$, and in the second and third columns when varying $\sum m_\nu$ and $N_\mathrm{eff}$, respectively.}\label{fig_dev}
\end{figure*}
\begin{figure*}
\centering
\includegraphics[scale=0.55]{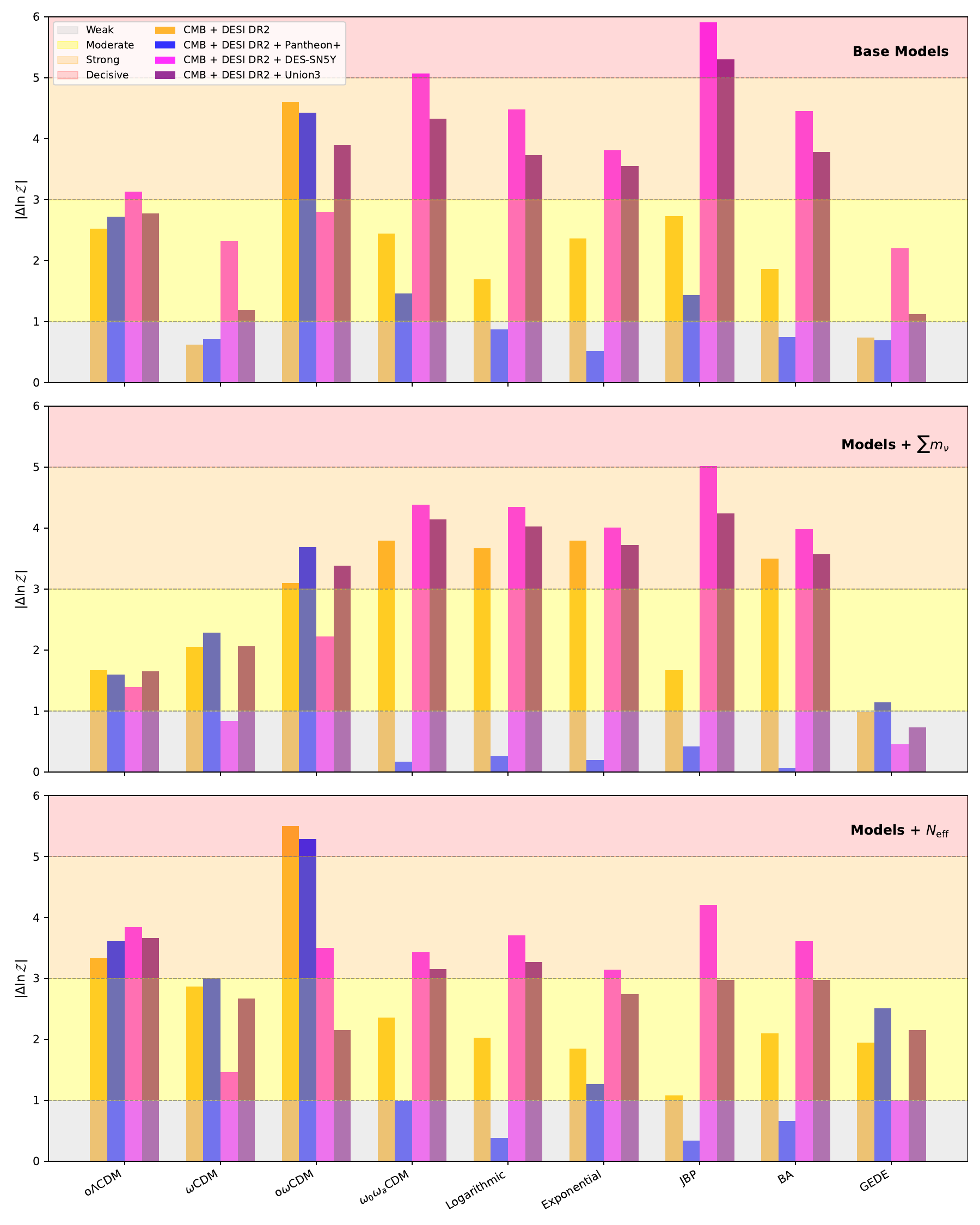}
\caption{The figure shows a comparison of various models relative to $\Lambda$CDM across different cosmological datasets, using the Bayesian evidence differences ($|\Delta \ln \mathcal{Z}|$). The first row corresponds to analyses with fixed $\sum m_{\nu}$ and $N_{\rm eff}$, the second row allows $\sum m_{\nu}$ to vary, and the third row allows $N_{\rm eff}$ to vary.)}\label{fig_bayes}
\end{figure*}
\section{Results}\label{sec_4}
In this section, we present strong evidence of dynamical DE. Tables~\ref{tab_2}, \ref{tab_3}, and \ref{tab_4} show the numerical values obtained from the MCMC analysis for each DE model. The results are presented in three distinct cases: first, without considering either $\sum m_\nu$ or $N_{\text{eff}}$, second, including $\sum m_\nu$, and finally, with $N_{\text{eff}}$.

Fig~\ref{fig_h} show the inferred values of inferred values of the \(h\) for each cosmological model. The first column shows to the baseline case without varying \(\sum m_\nu\) and \(N_\mathrm{eff}\); the second and third columns show the results when allowing \(\sum m_\nu\) and \(N_\mathrm{eff}\) to vary, respectively. The red bands shows the local measurement of \(h = 0.735 \pm 0.014\) \cite{riess2022comprehensive}, while the blue bands denote the \textit{Planck} 2018 constraint of \(h = 0.674 \pm 0.005\)~\cite{aghanim2020planck}. It also shows that none of the DE models predict a value of $h$ close to the Riess measurement, but they lie within the Planck precision. It is worth noting that in dynamical dark energy models where \( \omega > -1 \), the corresponding dark energy density scaling factor satisfies \( f_{\mathrm{DE}} > 1 \). Since dark energy primarily affects the late-time expansion, the combination \( \Omega_m h^2 \) remains approximately close to that predicted by the \(\Lambda\)CDM model. Consequently, if \( f_{\mathrm{DE}} \) increases, the only way to maintain consistency with the observed \( H(z) \) data is for the inferred value of \( H_0 \) to decrease, which in turn worsens the tension. This behavior has also been discussed in \cite{lee2022local,colgain2025much}. In \cite{colgain2025much}, it is also shown in Table~4 that the $\omega_0\omega_a$CDM model predicts $\omega_0 > -1$. The value of $H_0$ decreases accordingly when the DESI DR2 data are combined with the CMB and different SNe~Ia calibrations.

First, we compared the inferred values of the Hubble parameter $h$ and the matter density $\Omega_m$ for all DE parameterizations relative to the baseline $\Lambda$CDM model, using CMB+DESI~DR2 data both alone and combined with the Pantheon$^+$, DES-SN5Y, and Union3 Type~Ia supernova samples. Most models yield results consistent with $\Lambda$CDM within the $2\sigma$ level, with differences in both parameters typically smaller than $1\sigma$. Among the geometric extensions, o$\Lambda$CDM and $\omega$CDM predict the smallest deviations from the $\Lambda$CDM baseline, remaining below $1\sigma$ for both $h$ and $\Omega_m$. The o$\omega$CDM model also remains within $1\sigma$ deviation. In contrast, dynamically DE models, such as $\omega_0\omega_a$CDM, BA, JBP, Logarithmic, and Exponential, show deviations typically in the range of $1$–$2\sigma$. The Exponential model shows the largest deviations, reaching up to $\sim2.5\sigma$ in $h$ for some dataset combinations, while GEDE remains the closest to $\Lambda$CDM, showing deviation below $1\sigma$. Including additional degrees of freedom, such as the total neutrino mass $\sum m_\nu$ or the effective number of relativistic species $N_{\rm eff}$, does not show any significant deviation from the predicted values of $h$ and $\Omega_m$. The resulting estimates of $h$ and $\Omega_m$ remain compatible with $\Lambda$CDM within $2\sigma$, with only marginal shifts observed.

\paragraph{Without $\sum m_\nu$ and $N_{\text{eff}}$ :-} Fig.~\ref{fig_1} shows the parameter planes for different DE models, without the effects of $\sum m_\nu$ and $N_{\text{eff}}$. In particular, Figs.~\ref{fig_1a}, \ref{fig_1b}, and \ref{fig_1c} show the $\Omega_k$-$\Omega_m$, $\omega$-$\Omega_m$, and $\omega$-$\Omega_k$ planes for the o$\Lambda$CDM, $\omega$CDM, and o$\omega$CDM models, respectively. From Figs.~\ref{fig_1a} and \ref{fig_1c}, we observe that the preferred values of $\Omega_k$ are very close to zero, indicating consistency with a spatially flat Universe. In Figs.~\ref{fig_1b} and \ref{fig_1c}, we can also observe the preferred values of $\omega_0$. The CMB + DESI DR2 combination predicts a value close to the $\Lambda$CDM expectation of $\omega = -1$. However, when different calibrations for SNe Ia are included, the preferred value of $\omega_0$ begins to deviate from the $\Lambda$CDM model, shifting to $\omega_0 > -1$.

Figs.~\ref{fig_1d}, \ref{fig_1e}, \ref{fig_1f}, \ref{fig_1g}, and \ref{fig_1h} show the $\omega_0$-$\omega_a$ planes for the $\omega_0\omega_a$CDM, Logarithmic, Exponential, JBP, and BA models, respectively. In all cases, the preferred values of $\omega_0$ and $\omega_a$ deviate from the prediction of $\Lambda$CDM ($\omega_0 = -1$, $\omega_a = 0$). These deviations shows the evidence of a dynamical DE scenario, characterized by $\omega_0 > -1$, $\omega_a < 0$, and $\omega_0 + \omega_a < -1$, corresponding to a Quintom-B type behavior. Fig.~\ref{fig_1i} shows the $\Delta$-$\Omega_m$ plane for the GEDE model. It can be observed that the CMB + DESI DR2 combination predicts $\Delta > 0$, while the inclusion of different SNe~Ia calibrations preference $\Delta < 0$ which correspond to an earlier injection of DE.

\paragraph{Including $\sum m_\nu$ :-}
Fig.~\ref{fig_2} shows the parameter planes for each DE model, with $\sum m_\nu$ treated as a free parameter. In our analysis, we choose the prior $\sum m_\nu > 0$. Figs.~\ref{fig_2a}, \ref{fig_2b}, and \ref{fig_2c} show the $\Omega_k$-$\Omega_m$, $\omega$-$\Omega_m$, and $\omega$-$\Omega_k$ planes for the o$\Lambda$CDM, $\omega$CDM, and o$\omega$CDM models, respectively. The results for $\Omega_k$ and $\omega_0$ are consistent with the trends observed in Case 1, with the preferred values of $\Omega_k$ remaining close to zero, indicating a flat universe. Similarly, $\omega_0$ remains close to $-1$ for CMB + DESI~DR2 but shifts toward $\omega_0 > -1$ when different SNe~Ia calibrations are included.

Figs.~\ref{fig_2d}, \ref{fig_2e}, \ref{fig_2f}, \ref{fig_2g}, and \ref{fig_2h} show the $\omega_0$–$\omega_a$ planes for the $\omega_0\omega_a$CDM, Logarithmic, Exponential, JBP, and BA models, where similar deviations from the $\Lambda$CDM model are observed in case1, suggesting a dynamic DE scenario characterized by Quintom-B type behavior. Fig.~\ref{fig_2i} shows the $\Delta$-$\Omega_m$ plane for the GEDE model. The trend for $\Delta$ is similar to that seen in Case 1, with CMB + DESI DR2 predicting $\Delta > 0$, while the inclusion of various SNe~Ia calibrations prefers $\Delta < 0$, corresponding to an earlier injection of DE.

We also explore the upper bounds on the sum of neutrino masses within each DE model. First, in Fig.~\ref{fig_sum_LCDM}, we show the $\Omega_m$–$\sum m_\nu$ planes $\sum m_\nu$ for the $\Lambda$CDM model. Using only the CMB, we find $\sum m_\nu < 0.578$ eV, while combining CMB data with DESI DR2 we find $\sum m_\nu < 0.066$ eV. When combining CMB data with DESI DR2, the constraints become significantly tighter, as DESI favors lower $\Omega_m$ values. This constraint is reduced slightly to $< 0.073 \, \text{eV}$ when including the PP data set. The inclusion of the DES-SN5Y sample results in a marginally higher upper limit of $< 0.086 \, \text{eV}$, while the Union3 sample leads to a slightly tighter upper limit of $< 0.074 \, \text{eV}$. 

Fig.~\ref{fig_3} shows the parameter planes for models beyond the $\Lambda$CDM model. Fig.~\ref{fig_3a} shows the $\Omega_m$–$\sum m_\nu$ plane for the o$\Lambda$CDM model. Using the combination of CMB and DESI DR2 BAO data, we find a tight upper bound of $\sum m_\nu < 0.263,\text{eV}$, which loosens slightly to $< 0.267,\text{eV}$ when including the PP. Adding the DES-SN5Y sample further loosens the bound to $< 0.340,\text{eV}$, while the Union3 dataset slightly tightens it to $< 0.260,\text{eV}$.

Figs.~\ref{fig_3b} and \ref{fig_3c} show the $\Omega_m$–$\sum m_\nu$ planes for the $\omega$CDM and o$\omega$CDM models, respectively. It is worth noting that for the $\omega$CDM model, adding the DES-SN5Y and Union3 datasets to the CMB + DESI DR2 combination results in tightened upper bounds on $\sum m_\nu$ of $< 0.057,\text{eV}$ and $< 0.056,\text{eV}$, respectively. The bound loosens slightly to $< 0.062,\text{eV}$ when including the Pantheon$^+$ dataset, and with only CMB + DESI DR2, it is further relaxed to $< 0.075,\text{eV}$. In the case of the o$\omega$CDM model, the trend remains similar, but the bounds are generally weaker. Specifically, CMB + DESI DR2 + DES-SN5Y gives $\sum m_\nu < 0.197,\text{eV}$, CMB + DESI DR2 + Union3 gives $< 0.257,\text{eV}$, CMB + DESI DR2 + PP gives $< 0.260,\text{eV}$, and CMB + DESI DR2 alone gives $< 0.520,\text{eV}$.

Figs.~\ref{fig_3d}, ~\ref{fig_3e}, ~\ref{fig_3f},  ~\ref{fig_3g}, and  ~\ref{fig_3h}  show the $\omega_0\omega_a$–$\sum m_\nu$ planes for the $\omega_0\omega_a$CDM, Logarithmic, Exponential, JBP, and BA models. The gray dotted line at $\omega_0 = -1$ and $\omega_a = 0$ represents the $\Lambda$CDM prediction. In the case of the $\omega_0 \omega_a$CDM model, the tightest constraint on $\sum m_\nu$ is obtained when combining CMB + DESI DR2 + PP, giving $\sum m_\nu < 0.101,\text{eV}$. Adding DES-SN5Y slightly loosens the bound to $< 0.123,\text{eV}$, while including Union3 further relaxes the limit to $< 0.126,\text{eV}$. Using only CMB + DESI DR2 results in $< 0.127,\text{eV}$. Similarly, the Logarithmic model shows a constraint of $\sum m_\nu < 0.123,\text{eV}$ using CMB + DESI DR2 BAO data, which tightens to $< 0.067,\text{eV}$ when including the PP dataset. The limit relaxes slightly to $< 0.072,\text{eV}$ with DES-SN5Y and further to $< 0.085,\text{eV}$ when Union3 is added. In the Exponential model, the upper limit on $\sum m_\nu$ using CMB + DESI DR2 BAO data is $\sum m_\nu < 0.112,\text{eV}$. Adding the PP dataset tightens the constraint to $< 0.074,\text{eV}$, and including DES-SN5Y further tightens it to $< 0.067,\text{eV}$. Finally, adding Union3 slightly relaxes the upper limit to $< 0.077,\text{eV}$. In the case of the BA model, considering CMB + DESI DR2 alone gives an upper bound of $\sum m_\nu < 0.109,\text{eV}$. Adding the PP dataset tightens the constraint to $< 0.062,\text{eV}$, while including DES-SN5Y slightly loosens it to $< 0.072,\text{eV}$. Finally, adding Union3 further relaxes the upper limit to $< 0.078,\text{eV}$.

In the case of the JBP model, combining CMB + DESI DR2 yields a much tighter constraint than the $\omega_0\omega_a$CDM, Logarithmic, Exponential, and BA models when considering CMB + DESI DR2 alone, predicting an upper bound of $\sum m_\nu < 0.054,\text{eV}$. Within the JBP model, adding PP slightly relaxes the bound to $< 0.055,\text{eV}$, and including DES-SN5Y further loosens it to $< 0.059,\text{eV}$. Finally, adding Union3 tightens the constraint again to $< 0.056,\text{eV}$. Fig.~\ref{fig_3d} shows the $\Omega_m$–$\sum m_\nu$ parameter plane for the GEDE model. It is worth noting that in this model, combining CMB + DESI DR2 yields a much tighter constraint than the other models, with $\sum m_\nu < 0.043,\text{eV}$. Adding the PP dataset further tightens the constraint to $< 0.030,\text{eV}$, and including DES-SN5Y tightens it even more to $< 0.029,\text{eV}$. Finally, adding Union3 slightly relaxes the upper limit to $< 0.031,\text{eV}$. In Figs.~\ref{fig_posterior}, we also show the 1D marginalized posterior constraints on $\sum m_\nu$ using DESI DR2 + CMB. It can be observed that the GEDE and JBP models exhibit tighter constraints on $\sum m_\nu$ compared to the $\Lambda$CDM model.

\paragraph{Including $N_{\text{eff}}$ :-} 
Fig.~\ref{fig_4} shows the parameter planes of each DE model, including $N_{\text{eff}}$ as free parameters. Figs.~\ref{fig_4a}, \ref{fig_4b}, and \ref{fig_4c} show the $\Omega_k$-$\Omega_m$, $\omega$-$\Omega_m$, and $\omega$-$\Omega_k$ planes for the o$\Lambda$CDM, $\omega$CDM, and o$\omega$CDM models, respectively. The results for $\Omega_k$ and $\omega_0$ are consistent with the trends observed in the previous cases, with $\Omega_k$ staying near zero and $\omega_0$ shifting slightly toward $\omega_0 > -1$ when different SNe Ia calibrations are included. 

Figs.~\ref{fig_4d}, \ref{fig_4e}, \ref{fig_4f}, \ref{fig_4g}, and \ref{fig_4h} show the $\omega_0$-$\omega_a$ planes for the $\omega_0\omega_a$ CDM, logarithmic, exponential, JBP, and BA models, with deviations from the $\Lambda$CDM model once again pointing to the dynamical DE scenario, characterized by Quintom-B-type behavior. Finally, Fig.~\ref{fig_4i} presents the $\Delta$-$\Omega_m$ plane for the GEDE model. The results for $\Delta$ are consistent with the previous cases, showing a preference for positive $\Delta$ in the CMB + DESI DR2 combination, which shifts to negative values when different calibrations of SNe Ia are included. The 7th row of Table~\ref{tab_4} presents the constraints on the effective number of relativistic degrees of freedom. Our results are consistent with the standard particle physics value of $N_{\rm eff} = 3.044$, as reported in our supporting paper~\cite{elbers2025constraints}.

It is important to note that, the predicted values of $\Omega_k$, whether varying $\Sigma m_{\nu}$ or $N_{\mathrm{eff}}$, consistently indicate that $\Omega_k \approx 0 \;\Rightarrow\; k \approx 0$, i.e., the Universe is spatially flat. These predictions are in good agreement with earlier constraints obtained by WMAP ($-0.0179 < \Omega_k < 0.0081$, 95\% CL) \cite{hinshaw2009five}, BOOMERanG ($0.88 < \Omega_{M/R} + \Omega_\Lambda < 1.0081$, 95\% CL) \cite{de2000flat}, and Planck ($\Omega_{M/R} + \Omega_\Lambda = 1.00 \pm 0.026$, 68\% CL) \cite{aghanim2020planck}. This consistency is further shown in Fig.~\ref{fig_cmb}, where the geometry of hot and cold spots in the CMB also supports that overall curvature of the Universe is indeed flat (WMAP image source)\footnote{\url{https://map.gsfc.nasa.gov/media/030639/index.html}}. Further for the JBP parameterization, the predicted value of $\omega_a$ shifts toward significantly negative values, extending beyond previous limits and allowing for $\omega_0 > -1$.

In Fig.~\ref{fig_dev}, we also show the deviation from the $\Lambda$CDM model without varying neutrino mass or effective neutrino number in the first column. The deviation from the $\Lambda$CDM model depends on the choice of SNe~Ia sample combined with CMB + DESI DR2. With Pantheon$^{+}$, most models show only mild to moderate deviations at the $1\sigma$–$2.3\sigma$ level, suggesting broad consistency with $\Lambda$CDM. In contrast, the inclusion of DES-SN5Y significantly increases the deviations: models such as o$\omega$CDM, $\omega_0\omega_a$CDM, BA, and GEDE show tensions exceeding $3\sigma$, pointing toward possible departures from a pure cosmological constant. A similar effect is seen with Union3, where $\omega_0\omega_a$CDM, Exponential, BA, and JBP all remain in the $3\sigma$ range, strengthening the evidence for dynamical DE.

The second column of Fig.~\ref{fig_dev} shows the deviations from $\Lambda$CDM when the total neutrino mass $\sum m_\nu$ is allowed to vary. With CMB + DESI DR2 alone, most models stay within $2.5\sigma$, though $\omega_0\omega_a$CDM shows a stronger $3.4\sigma$ deviation. Adding Pantheon$^{+}$ reduces the tensions overall, keeping most models below $2.5\sigma$. In contrast, including DES-SN5Y significantly increases the deviations, with $\omega_0\omega_a$CDM, Logarithmic, Exponential, and JBP models all exceeding $3\sigma$, reaching nearly $3.9\sigma$ in the case of $\omega_0\omega_a$CDM. A similar trend appears with Union3, where multiple models (e.g., $\omega_0\omega_a$CDM, Exponential, BA, JBP) remain above $3\sigma$

The third column of Fig.~\ref{fig_dev} shows the deviations from $\Lambda$CDM when the effective number of neutrino species, $N_{\rm eff}$, is allowed to vary. The deviations from $\Lambda$CDM show a clear dependence on the choice of datasets. With CMB + DESI DR2 alone, most models remain below $2.8\sigma$, with BA showing the largest tension at $2.77\sigma$. Adding Pantheon$^+$ reduces the deviations, keeping nearly all models below $2.5\sigma$ and closer to $\Lambda$CDM. Including DES-SN5Y significantly increases the tensions, with $\omega_0 \omega_a$CDM, BA, and JBP models exceeding $3\sigma$. A similar trend is observed with Union3, where $\omega_0 \omega_a$CDM, BA, and JBP remain above $3\sigma$.

In Fig.~\ref{fig_bayes}, we show  the comparative analysis of different cosmological models relative to the baseline $\Lambda$CDM model, using Bayesian evidence, without varying $\sum m_\nu$ and $N_{\rm eff}$ (1st row). When considering the CMB + DESI DR2 dataset, the o$\omega$CDM model shows the strongest preference with $|\Delta \ln \mathcal{Z}| = 4.60$, indicating strong evidence against $\Lambda$CDM, while the $\omega$CDM and GEDE models are weakly disfavored ($|\Delta \ln \mathcal{Z}| = 0.62$ and 0.74, respectively), and the remaining models show moderate evidence. When adding Pantheon$^+$ SNe Ia data, the o$\omega$CDM model remains moderately disfavored ($|\Delta \ln \mathcal{Z}| = 4.43$), whereas most other models exhibit weak or inconclusive evidence relative to $\Lambda$CDM ($|\Delta \ln \mathcal{Z}| < 1.5$). For CMB + DESI DR2 + DES-SN5Y, the JBP model shows decisive evidence against $\Lambda$CDM ($|\Delta \ln \mathcal{Z}| = 5.91$), followed by the $\omega_a \omega_0$CDM (5.07) and Logarithmic (4.48) models with strong evidence, while the GEDE model remains only weakly disfavored (2.20). Finally, with CMB + DESI DR2 + Union3, the JBP model again shows strong preference against $\Lambda$CDM ($|\Delta \ln \mathcal{Z}| = 5.30$), with o$\omega$CDM and $\omega_a \omega_0$CDM also moderately disfavored (3.90 and 4.33, respectively), and models such as $\omega$CDM and GEDE remaining largely consistent with $\Lambda$CDM ($|\Delta \ln \mathcal{Z}| \lesssim 1.2$).

The 2nd column of Fig.~\ref{fig_bayes} shows that when allowing the sum of neutrino masses, $\sum m_\nu$, to vary, the Bayesian evidence for different models relative to $\Lambda$CDM + $\sum m_\nu$ shows noticeable shifts. For the CMB + DESI DR2 dataset, the $\omega_0 \omega_a$CDM, Logarithmic, and Exponential models are moderately to strongly disfavored relative to $\Lambda$CDM + $\sum m_\nu$ ($|\Delta \ln \mathcal{Z}| = 3.79, 3.67, 3.79$, respectively), while o$\omega$CDM and BA models exhibit moderate evidence against $\Lambda$CDM + $\sum m_\nu$ (3.10 and 3.50). In contrast, o$\Lambda$CDM, JBP, and GEDE models remain weakly disfavored ($|\Delta \ln \mathcal{Z}| \lesssim 1.0$). Including Pantheon$^+$ SNe Ia data generally weakens the evidence, with most models showing weak or inconclusive preference ($|\Delta \ln \mathcal{Z}| \lesssim 0.5$), except for o$\omega$CDM and $\omega$CDM which remain moderately disfavored (3.69 and 2.28). For CMB + DESI DR2 + DES-SN5Y, the JBP model displays decisive evidence against $\Lambda$CDM + $\sum m_\nu$ ($|\Delta \ln \mathcal{Z}| = 5.02$), while $\omega_0 \omega_a$CDM, Logarithmic, Exponential, and BA models are strongly disfavored (4.38, 4.35, 4.01, 3.98), and the remaining models are only weakly disfavored. Finally, for CMB + DESI DR2 + Union3, the $\omega_0 \omega_a$CDM, Logarithmic, Exponential, JBP, and BA models are strongly to moderately disfavored ($|\Delta \ln \mathcal{Z}|$ between 3.57 and 4.24), while o$\Lambda$CDM, $\omega$CDM, and GEDE remain largely consistent with $\Lambda$CDM + $\sum m_\nu$ ($|\Delta \ln \mathcal{Z}| \lesssim 2.1$).

The 3rd column of Fig.~\ref{fig_bayes} shows that when allowing the effective number of relativistic species, $N_{\rm eff}$, to vary, the Bayesian evidence for different models relative to $\Lambda$CDM + $N_{\rm eff}$ exhibits notable differences. For the CMB + DESI DR2 dataset, the o$\omega$CDM model is strongly disfavored ($|\Delta \ln \mathcal{Z}| = 5.50$), followed by o$\Lambda$CDM and $\omega$CDM with moderate to strong evidence against $\Lambda$CDM + $N_{\rm eff}$ ($|\Delta \ln \mathcal{Z}| = 3.33$ and 2.87, respectively), while the remaining models are only weakly disfavored ($|\Delta \ln \mathcal{Z}| \lesssim 2.1$). Adding Pantheon$^+$ SNe Ia data generally weakens the evidence, with most models showing weak or inconclusive preference ($|\Delta \ln \mathcal{Z}| \lesssim 1.3$), except o$\omega$CDM and o$\Lambda$CDM which remain strongly disfavored (5.29 and 3.62). For CMB + DESI DR2 + DES-SN5Y, JBP shows decisive evidence against $\Lambda$CDM + $N_{\rm eff}$ ($|\Delta \ln \mathcal{Z}| = 4.21$), with $\omega_0 \omega_a$CDM, Logarithmic, Exponential, BA, and o$\Lambda$CDM models being strongly disfavored ($|\Delta \ln \mathcal{Z}|$ between 3.14 and 3.84), while $\omega$CDM and GEDE remain weakly disfavored. Finally, for CMB + DESI DR2 + Union3, multiple models, including $\omega_0 \omega_a$CDM, Logarithmic, Exponential, JBP, and BA, show moderate evidence against $\Lambda$CDM + $N_{\rm eff}$ ($|\Delta \ln \mathcal{Z}|$ between 2.15 and 3.27), whereas $\omega$CDM and GEDE remain mostly consistent with the base model ($|\Delta \ln \mathcal{Z}| \lesssim 2.15$).

The seventh rows of Tables~\ref{tab_2},~\ref{tab_3}, and~\ref{tab_4} present the $\Delta\chi^2$ values obtained when the effects of $\sum m_\nu$ and $N_{\text{eff}}$ are neglected, when $\sum m_\nu$ is allowed to vary, and when $N_{\text{eff}}$ is allowed to vary, respectively. In all three cases, the dynamical DE models particularly the $\omega_0\omega_a$CDM, Logarithmic, Exponential, JBP, and BA models consistently yield lower $\chi^2$ values compared to $\Lambda$CDM, with improvements reaching up to $\Delta\chi^2 \simeq -7$ for the combined CMB + DESI DR2 + DES-SN5Y dataset. In contrast, for some combinations of the datasets, the GEDE model shows positive $\Delta\chi^2$ values, indicating a less favorable fit relative to $\Lambda$CDM.

In the context of BAO DESI DR2, \cite{colgain2025much} found that LRG2 has a stronger impact on the dynamical DE signal than LRG1. This is shown in Figure 4 of \cite{colgain2025much}, where the faded blue curve represents the DR1 data. In DESI DR1, the dynamical DE signal is influenced by a combination of high $\Omega_m$ (LRG1), low $\Omega_m$ (LRG2), and higher $\Omega_m$ at higher redshifts, with a similar pattern observed in DESI DR2, but shifted to higher redshifts. Further details on LRG2 are provided in the appendix of \cite{goldstein2025monodromic}. Moreover, \citep{Huang2025LowZBias, gialamas2025interpreting,chaudhary2025evidence} show that the dynamical DE behavior is primarily driven by low-redshift supernovae ($z < 0.1$). When these supernovae are excluded from the analysis, the results return to the standard $\Lambda$CDM model. Additionally, \citep{Huang2025LowZBias} highlights potential inconsistencies within the DES-SN5Y compilation, which appears less self-consistent than the Pantheon$^+$ dataset. Figs.~2 in \cite{lodha2025extended,leauthaud2025looking} and Fig.~10 in \cite{chaudhary2025evidence} shows the corresponding evolution of the EoS parameter. These findings suggest that $\omega$ is not constant: DE enters a phantom phase ($\omega < -1$) at $z > 0.5$, crosses the phantom divide around $z \sim 0.5$, and evolves rapidly at $z < 1$, eventually reaching $\omega > -1$. Establishing $\omega \neq -1$ would represent a major outcome.

The $(\omega_0, \omega_a)$ are determined within a finite redshift interval, and extending them to all epochs does not always reflect the actual physical evolution. Consequently, even non-phantom models can yield best-fit $(\omega_0, \omega_a)$ values that appear to cross $\omega = -1$ when the CPL form is extrapolated beyond the fitted range. This apparent crossing arises from the limits of the parameterization rather than from a genuine phantom transition~\cite{wolf2023underdetermination,shlivko2024assessing,wolf2024scant,cortes2024interpreting}. Also, the regions of the $(\omega_0, \omega_a)$ plane associated with apparent phantom behavior are not limited to multi-field Quintom constructions. Recent work shows that single-field DE models with non-minimal couplings to gravity can also produce stable phantom-like behavior and populate similar regions favored by current data~\cite{wolf2025assessing,wolf2025matching,adam2025comparing}.

There are several other interesting findings from DESI DR2, an interesting finding from DESI DR2 is related to the ongoing $H_0$ tension. Although dynamical DE can explain the BAO + CMB + SNe data, it does not solve the $H_0$ tension. In fact, when combining these data with local measurements of $H_0$, dynamical DE actually makes the disagreement worse. Specifically, in the $\Lambda$CDM model, the DESI + Planck data produce $H_0 = 68.17 \pm 0.28$ km s$^{-1}$ Mpc$^{-1}$, corresponding to a 5$\sigma$ tension. The best-fit dynamical DE models predict $H_0$ values ranging from $63.6^{+1.6}_{-2.1}$ to $67.51 \pm 0.59$ km s$^{-1}$ Mpc$^{-1}$, depending on the assumptions used. None of these values match the broader range of precise local $H_0$ measurements, which highlights the ongoing tension in the $H_0$ sector, even with dynamical DE included \cite{verde2024tale}.
\section{Conclusions and Final Remarks}\label{sec_5}
In this work, we have examined strong evidence for dynamical DE by exploring several DE models beyond the standard $\Lambda$CDM model. Using the standard FLRW metric and assuming a spatially flat Universe, we derived the dimensionless expansion function and analyzed different classes of DE models, from the cosmological constant to time-varying EoS models such as $\omega_0\omega_a$CDM, Logarithmic, Exponential, JBP, BA, and GEDE. We also considered non flat models to account for the possible spatial curvature of the Universe. These models were then tested against a variety of observational data to constrain crucial cosmological parameters.

At the observational data level, we performed Bayesian parameter estimation using the \texttt{SimpleMC} cosmological inference code, integrated with the Metropolis-Hastings MCMC algorithm. We utilized data from Baryon Acoustic Oscillations BAO from DESI DR2, Type Ia Supernovae, and Cosmic Microwave Background distance priors to constrain the parameters of each cosmological model. Additionally, we extended the analysis to include the sum of neutrino masses ($\sum m_\nu$) and the effective number of relativistic species ($N_{\text{eff}}$), allowing us to investigate their impact on cosmological parameters and derive tighter constraints on the models.

Through our analysis, we found several key results:

\begin{itemize}
    \item \textbf{Dynamical Dark Energy}: Our analysis suggests that $\omega \neq -1$, as indicated by the preferred deviations from the $\Lambda$CDM model. Specifically, the models deviate from the point ($\omega_0 = -1$, $\omega_a = 0$) and prefer values such that $\omega_0 > -1$, $\omega_a < 0$, and $\omega_0 + \omega_a < -1$, pointing to a dynamical DE scenario. This is especially true for models such as $\omega_0\omega_a$ CDM, logarithmic, exponential, and JBP, where the parameters $\omega_0$ and $\omega_a$ show signs of dynamical behavior, indicating Quintom-B-type behavior. These findings show strong evidence of dynamical DE
    
    \item \textbf{Spatial Curvature}: The inclusion of spatial curvature, particularly in the $\Lambda$CDM and $\omega$CDM models, did not significantly alter the cosmological parameters derived from the data. The preferred value for the curvature parameter, $\Omega_k$, remained close to zero, suggesting that the universe is still largely consistent with being spatially flat, as observed in the CMB data.
    
    \item \textbf{Neutrino Masses}: We derived upper bounds on the total neutrino mass $\sum m_\nu$ and found that combining CMB and DESI DR2 data with SNe Ia datasets provides strong constraints across all DE models. Extended models such as JBP and GEDE yield the tightest limits, $\sum m_\nu \sim 0.03$–0.056 eV, while models like o$\omega$CDM give comparatively weaker bounds. These results demonstrate that both the choice of DE model and the dataset combination significantly affect neutrino mass constraints, with extended DE models allowing notably tighter limits.
    
    \item \textbf{Effective Relativistic Species}: For the effective number of relativistic species, our results are consistent with the standard value of $N_{\text{eff}} = 3.044$, as expected from particle physics. However, the inclusion of different SNe Ia data sets and the consideration of spatial curvature had a small influence on the $N_{\text{eff}}$ parameter, with values still consistent with the standard model.
    
    \item \textbf{$5\sigma$ deviation from $\Lambda$CDM} None of the models reach the $5\sigma$ threshold of deviation from $\Lambda$CDM. While certain dynamical DE models particularly $\omega_0\omega_a$CDM, BA, JBP, and Exponential show tensions exceeding $3\sigma$ with DES-SN5Y or Union3, the deviations remain below the $5\sigma$ discovery level. This indicates that, even allowing for variable neutrino mass or $N_{\rm eff}$, $\Lambda$CDM remains broadly consistent with current datasets, while hints of dynamical DE appear, but it is still too early to confirm it
    
    \item \textbf{Statistical Significance:} Our analysis shows that extended models with dynamic DE or curvature are generally preferred over $\Lambda$CDM, with moderate to strong evidence when SNe Ia datasets such as DES-SN5Y and Union3 are included; allowing $\sum m_\nu$ to vary slightly reduces the statistical tension for some models, particularly with Pantheon$^+$ data, while varying $N_{\rm eff}$ further emphasizes the preference for curvature or dynamical DE models, especially for datasets including DES-SN5Y. The $\Delta\chi^2$ analysis across all scenarios shows that dynamical DE models consistently achieve a better fit than the $\Lambda$CDM Model.
\end{itemize}

Through our work, we have shown that $\Lambda$CDM is challenged by the DESI DR2 data, which shows that DE is evolving. Although it is too early to draw definitive conclusions, the signs of cracks in the $\Lambda$CDM model are becoming apparent. If these deviations from $\Lambda$CDM are confirmed with a 5$\sigma$ certainty in the near future, it would mark a significant shift in cosmology, reminiscent of past paradigm changes that were accompanied by considerable conflict and resistance within the scientific community. The potential breakdown of the $\Lambda$CDM model raises the question of what we can learn from this transition. While we could discover a new, elegant theory as simple and transformative as General Relativity, it is equally possible that the new model will be more complex, incorporating a variety of components that challenge our current understanding. This could include multiple forms of dark matter, diverse DE fields with different properties, and even interactions between these components that affect the expansion of the Universe. Moreover, a breakdown of General Relativity on cosmological scales could introduce additional layers of complexity. In this scenario, we may be far from a clear and unified physical model, and the search for a complete understanding of the dark sector could stretch for many years. Cosmology has traditionally sought simplicity, often inspired by high-energy physics, but if the dark sector proves to be as intricate as the luminous sector, our assumptions about simplicity may no longer be valid. As we move forward, we should remain guided by increasingly robust observational data, secured through substantial investment, in our pursuit of a deeper understanding of the Universe.

\section*{\textbf{Appendix: Can Dynamical Dark Energy Model Solve the $H_0$ Tension ?}}
\begin{figure}[H]
\centering
\includegraphics[scale=0.56]{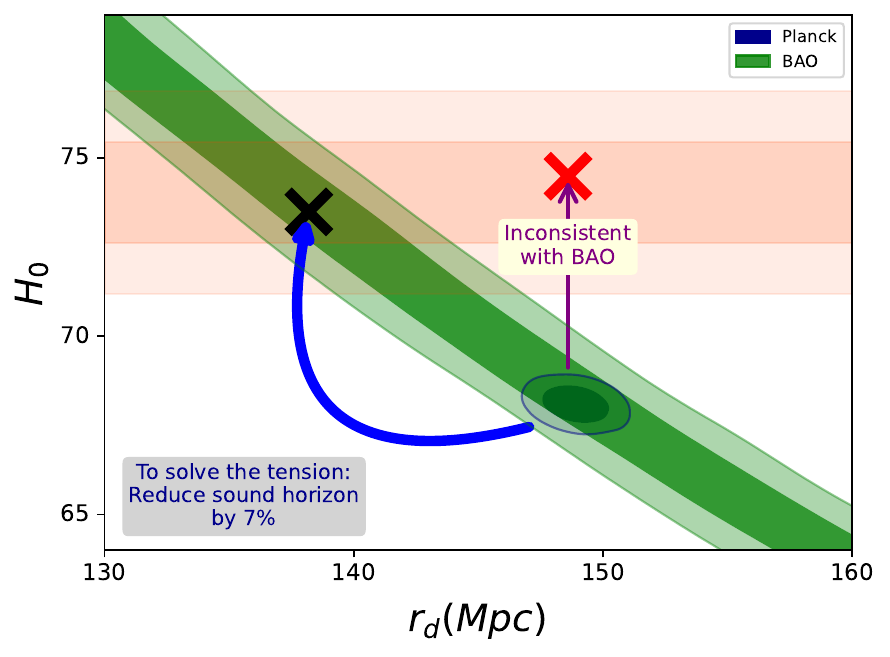}
\caption{The figure shows the $H_0 - r_d$ plane, where the blue contours represent the Planck predictions, the green bar shows the BAO measurements, and the light orange region represents the Riess measurements.}\label{fig_5}
\end{figure}

We would like to highlight an important point regarding the Dynamical DE model. When we consider Dynamical DE to explain the Universe, it becomes very difficult to resolve the Hubble tension. This can be understood in Fig.~\ref{fig_5}, where we show that in order to solve the Hubble tension, the Planck measurements should align with the black cross. In fact, if the value of $H_0$ increases, the sound horizon should decrease by approximately 7\%. However, in the case of the Dynamical DE model, this is not possible because the Dynamical DE model cannot lower the sound horizon, as DE is completely sub-dominant at recombination. In fact, even if one increases $H_0$, one would end up with the red cross, which is inconsistent with the BAO measurements.
\section*{Acknowledgements}
SC acknowledges the Istituto Nazionale di Fisica Nucleare (INFN) Sez. di Napoli,  Iniziative Specifiche QGSKY and MoonLight-2  and the Istituto Nazionale di Alta Matematica (INdAM), gruppo GNFM, for the support.
This paper is based upon work from COST Action CA21136 -- Addressing observational tensions in cosmology with systematics and fundamental physics (CosmoVerse), supported by COST (European Cooperation in Science and Technology).

\bibliographystyle{elsarticle-num}
\bibliography{mybib.bib}

\end{document}